\documentclass[pre,twocolumn,aps,showpacs,superscriptaddress,longbibliography]{revtex4-2}

\usepackage[T1]{fontenc}
\usepackage[utf8]{inputenc}
\usepackage{babel}
\usepackage{calc}
\usepackage{amsmath,bm,amssymb}

\usepackage{microtype}
\usepackage{lmodern}
\usepackage{graphicx}
\usepackage{float}
\usepackage{color, xcolor}
\definecolor{dkgreen}{rgb}{0.0,0.6,0.0} 
\definecolor{blue}{rgb}{1.0,0.49,0.0}
\usepackage[unicode=true,pdfusetitle,
 bookmarks=true,bookmarksnumbered=false,bookmarksopen=false,
 breaklinks=false,pdfborder={0 0 1},backref=false,colorlinks=true]
 {hyperref}
\hypersetup{citecolor=dkgreen,linkcolor=blue,urlcolor=blue}

\begin{document}

\title{
Phase space geometry of collective spin systems: Scaling and Fractality 
}

\author{Miguel Gonzalez}
\affiliation{Instituto de Ciencias Nucleares, Universidad Nacional Aut\'onoma de M\'exico, Apdo. Postal 70-543, C.P. 04510  Cd. Mx., Mexico}
\author{Miguel A.~Bastarrachea-Magnani}
\affiliation{Departamento de F\'isica, Universidad Aut\'onoma Metropolitana-Iztapalapa, Av. Ferrocarril San Rafael Atlixco 186, C.P. 09310 Mexico City, Mexico}
\author{Jorge G. Hirsch} 
\affiliation{Instituto de Ciencias Nucleares, Universidad Nacional Aut\'onoma de M\'exico, Apdo. Postal 70-543, C.P. 04510  Cd. Mx., Mexico}

\date{\today}

\begin{abstract}
We examine the scaling of the inverse participation ratio of spin coherent states in the energy basis of three collective spin systems: a bounded harmonic oscillator, the Lipkin-Meshkov-Glick model, and the Quantum Kicked Top. The finite-size quantum probing provides detailed insights into the structure of the phase space, particularly the relationship between critical points in classical dynamics and their quantum counterparts in collective spin systems. We 
introduce a finite-size scaling mass exponent that makes it possible to identify conditions under which a power-law behavior emerges, allowing to assign a fractal dimension to a coherent state. 
For the Quantum Kicked Top, the fractal dimension of coherent states -when well-defined- exhibits three general behaviors: one related to the presence of critical points and two associated with regular and chaotic dynamics. The finite-size scaling analysis paves the way toward exploring collective spin systems relevant to quantum technologies within the quantum-classical framework. 
\end{abstract}

\maketitle

\section{\label{sec:int} Introduction}

Phase space methods are becoming a valuable tool for
characterizing quantum systems with finite-dimensional Hilbert spaces ~\cite{phasespacemethod1}, particularly for systems with few degrees of freedom.
They also play a central role in studying the quantum chaos of systems with a well-defined (semi)classical limit, as is the case for collective spin systems~\cite{suncoherentstates1}, where spin coherent states are the quantum states used to construct such phase space. It is then an essential task to characterize the behavior of the spin coherent states expressed in the eigenbasis of a particular collective spin system. Also, the spin coherent states are central in addressing several questions about quantum chaos in the context of the quantum-classical correspondence. To investigate the latter, one has to increase the dimension of the Hilbert space to make the relevant Planck constant tend to zero, and this is how the fundamental question of this work about the spin coherent states arises: how do the spin coherent states scale in a given eigenbasis, as a function of the dimension of the Hilbert space?

The present contribution is devoted to analyzing the geometry of the phase space of quantum collective spin models (or single $\text{J}$ spin models~\cite{phasespacemethod1}) given by the structure of the eigenvectors of the particular Hilbert space in the context of a classical-quantum correspondence. To do so, we study and classify the scaling of the general inverse participation ratio ($\text{IPR}$) of spin coherent states on different eigenbasis, according to their location on the Poincar\'e surface of section. First, we exhibit the importance of studying the behavior of the scaling, as it is not guaranteed that a power-law behavior develops to ensure a fractal analysis. We demonstrate the $\text{IPR}$ scaling possesses three overall behaviors offering different information: the first one is related to its closeness to critical points; the second signals the regular or chaotic nature of the region where it is located; and the third one indicates specific dynamics around the critical point where the spin coherent state is located, which will be addressed in a future work~\cite{Miguel2025}. 

The scaling of the moments of the distribution of some quantity can have a critical exponent associated with its fractal dimension or an infinite set of critical exponents associated with multifractality~\cite{Mandelbrot1974,Ott2002}. In quantum mechanics, it characterizes the extension of a state in a given basis of the Hilbert space \cite{Mirlin2000,Nakayama2013}.
There is a close relationship between the concepts of localization in the classical phase space and its quantum counterpart in the Hilbert space \cite{Bastarrachea2016, Haake2018,Santhanam2022}. In the present contribution, these concepts are explored employing the scaling of the $\text{IPR}$ and, when available, its fractal dimension. 

The structure of this article is as follows. In Sec~\ref{sec:2}, we define the tools for a finite-size scaling analysis, including the inverse participation ratio ($\text{IPR}$), its generalized version ($\text{IPR}_{q}$), and introduce the concept of the finite-size mass exponent, which allows for both a fractal analysis of the system and a quantitative characterization of the scaling behavior of spin coherent states. 
We summarize the quantum-classical correspondence of collective spin Hamiltonians via spin coherent states in Sec.~\ref{sec:3}. In Sec.~\ref{sec:4}, we study the scaling of spin coherent states in the Dicke basis to benchmark the typical behavior of regular dynamics and show indicators of the non-power law behavior around fixed points. 
The following sections present
the finite-size scaling analysis in paradigmatic models. In Sec.~\ref{sec:5}, we study the scaling of the $\text{IPR}$ in the eigenbasis of the Lipkin-Meshkov-Glick (LMG) model, which stands as a point of reference for only regular dynamics and unstable fixed points. In Sec.~\ref{sec:6}, we perform the same analysis applied to the Quantum Kicked Top (QKT) for spin coherent states in the Floquet basis, exhibiting that it has three different (mono)fractal regimes, as well as regions where the asymptotic limit has not been reached. No fractal study can be performed there, but the finite-size analysis still contains valuable information about phase space geometry. Finally, in Sec.~\ref{sec:7}, we present our conclusions.

\section{Inverse Participation Ratio and fractality}
\label{sec:2}

The inverse participation ratio $\text{IPR}$ is a central tool to study the localization of a quantum state $|\psi\rangle$~\cite{localization1,Bell1970,Thouless1974}. For a given basis $\{|\phi_{i}\rangle\}_{i=1}^{N}$ with finite dimension $N$, the $\text{IPR}$ takes values between $1/N$ and $1$, for a maximally delocalized state and a maximally localized state, respectively. Here, we employ the generalized inverse participation ratio $\text{IPR}_{q}$, defined as~\cite{anderson1,Evers2008}
\begin{equation}
    \text{IPR}_{q} := \sum_{i}^{N} p_{i}^q \, ,
    \label{eqn:iprq}
\end{equation}
where the sum runs over all the elements of the basis under consideration, and the coefficients $p_{i} = |c_{i}|^{2} = |\langle \phi_{i}|\psi\rangle|^{2}$ are interpreted as a probability distribution due to the normalization condition $\sum_{i}p_{i}=1$. This quantity depends on the particular state under consideration and the basis on which it is expressed. The case $q = 1$ is the state's normalization condition, while $q=2$ coincides with the usual definition of the $\text{IPR}$. The $\text{IPR}_{q}$ in Eq.~\ref{eqn:iprq} allows us to obtain a quantitative picture of how the state $|\psi\rangle$ distributes over the basis under consideration at different scales, controlled by the parameter $q$, as a function of the basis dimension. The effect of $q$ in the range $0<q<1$ is to amplify the contribution of the smallest probabilities $p_{i}$: the closer $q$ is to zero, the larger the amplification. Instead, in the range $q>1$, the effect emphasizes the intensity of the largest probabilities in the distribution. In the following, we will restrict to the interval $q\in(0,4]$ to avoid numerical difficulties~\cite{Chen_2004}.

The scaling of $\text{IPR}_{q}$ as a function of $q$ has become a customary tool for multifractal analysis in several contexts~\cite{Ott2002,anderson1} from turbulent flows~\cite{Mandelbrot1974} and strange attractors~\cite{Grassberger1983a,Grassberger1984}, to a vast array of phenomena in quantum systems~\cite{Mirlin2006,Rodriguez2011,Roy2018,Lindinger2019,Solorzano2021,Sarkar2021,Sarkar2022}. However, a necessary condition to perform the formal multifractal analysis of a quantum state is that, for each value of $q$, the $\text{IPR}_{q}(N)$ behaves asymptotically as a power-law for large $N$~\cite{BKramer_1993, PhysRevLett.62.1327, PhysRevLett.122.106603, PhysRevB.57.10240, floquetfinetuning1,Bastarrachea2024}. In that case, the scaling of the $\text{IPR}_{q}$ signals the absence of length scales~\cite{PhysRevB.78.195106}, a concept related to criticality. 

Although there are examples in which multifractality for a single state can be analytically described, as, for example, spin coherent states in the spin-z (or computational basis) basis~\cite{spinchains1}, in most quantum systems, it has to be done numerically. This presents a challenge to study the scaling in quantum systems, not only due to computational limitations but also because, in general, there is no simple correspondence for eigenvectors calculated for different dimensions \cite{eigenamplitudes1}. Only a few states can be easily identified at any dimension, e.g., the ground-state and the highest energy state of a time-independent Hamiltonian in a finite-dimensional Hilbert space. This difficulty increases when one considers time-dependent, periodic Hamiltonians, as there is neither a ground nor maximum energy state. There is a work in progress~\cite{ipralgorithm1} in which it is proposed a novel way to directly estimate the \text{IPR} of a quantum state.

In the present work, we investigate the scaling of coherent states $|z(Q,P)\rangle$ distributed in three different eigenbasis basis $\{|\phi_{i}\rangle\}_{i=1}^{N}$, where $N=2J+1$. Here, the coefficients squared: $|\langle \phi_{i}|z(Q,P) \rangle|^2$ have a simple and valuable geometrical interpretation: they are the Husimi distribution of the Eigenvectors in phase space. The Husimi function can represent the structures observed in the Poincaré section in the semiclassical description, a phenomenon known as the principle of uniform semiclassical condensation of Wigner functions of eigenstates (PUSC) \cite{pusc1, powerlaymixed1}. In a mixed phase space, where regular and chaotic dynamics coexist, there are Husimi distribution of some eigenvectors that cover both regions in the classical phase space, whose study is still in progress
\cite{floodingstates1, powerlaymixed1}. 
When $J$ increases, approaching the classical limit, the coherent states remain centered at the same point in phase space, with its area diminishing as $1/J$, allowing the study of its scaling as function of $J$. 

When the $\text{IPR}_{q}$ scales as a power-law for large $N$, one can employ the mass exponents as a tool to perform a multifractal analysis~\cite{Ott2002,Bastarrachea2024}. They read
\begin{equation}
    \tau_{q} := -\lim_{N\to \infty} \dfrac{\ln{\text{IPR}_{q}}}{\ln{N}} \, 
    \label{eqn:tauq}
\end{equation}
and are calculated 
by
linearly fitting the ratio between $\ln\text{IPR}$ and $\ln{N}$. Moreover, one can write the mass exponents as 
\begin{equation}
\tau_{q}=D_{q}(q-1),
\label{eqn:Dq}
\end{equation}
where $D_{q}$ is called is the generalized (R\'enyi) dimension~\cite{Hentschel1983,Halsey1986}. When $D_{q}$ is constant, we talk about a monofractal with Hausdorff dimension $D_{0}$. Instead, when $D_{q}$ is a function of $q$, hence $\tau_{q}$ being non-linear, we call the object multifractal~\cite{Ott2002}. For the particular case of $q=2$, corresponding to the standard $\text{IPR}$ it becomes $\tau_{2}=D_{2}$, so the mass exponent is equal to the generalized dimension, and $D_{2}$ is also called correlation dimension~\cite{Grassberger1983a,Grassberger1984}.

Here, we generalize this definition to explore the scaling at finite values of $N$. We define the {\it finite mass exponent} $\tau$ as the slope of the $\ln\text{IPR}$ vs $\ln N$ 
between two consecutive points of the $\text{IPR}$ curve.
\begin{eqnarray}
    \tau := 
    \dfrac{\ln(\text{IPR}(N_2)/\text{IPR}(N_1))}{\ln(N_{2}/N_{1})}\, .
    \label{eqn:massexp}
\end{eqnarray}
For 
spin coherent states, we have analyzed different values of $q$, and concluded that all values provide the same information as $q=2$, the case in which we concentrate in the present study.

Via the finite mass exponent, we 
identify and characterize changes in the behavior of $\text{IPR}$ as a function of $N$. It is worth noting that it is not aimed to measure the number of basis states needed to express the state of interest, as it does, e.g., the number of elements to fulfill the normalization condition  $\text{IPR}_{1}=\sum_{i}p_{i}=1$. Instead, it provides information on the ratio between this number and the Hilbert space dimension. Furthermore, it has no restrictions to its functional behavior as it does the standard mass exponent $\tau_{q}$, whose curvature must be positive for all $q>0$~\cite{Halsey1986,Janssen1994,Ott2002}. It only serves as a quantitative measure to analyse the changes as the dimension of the Hilbert space varies.

\begin{figure}[t]
    \centering \includegraphics[width=1\columnwidth]{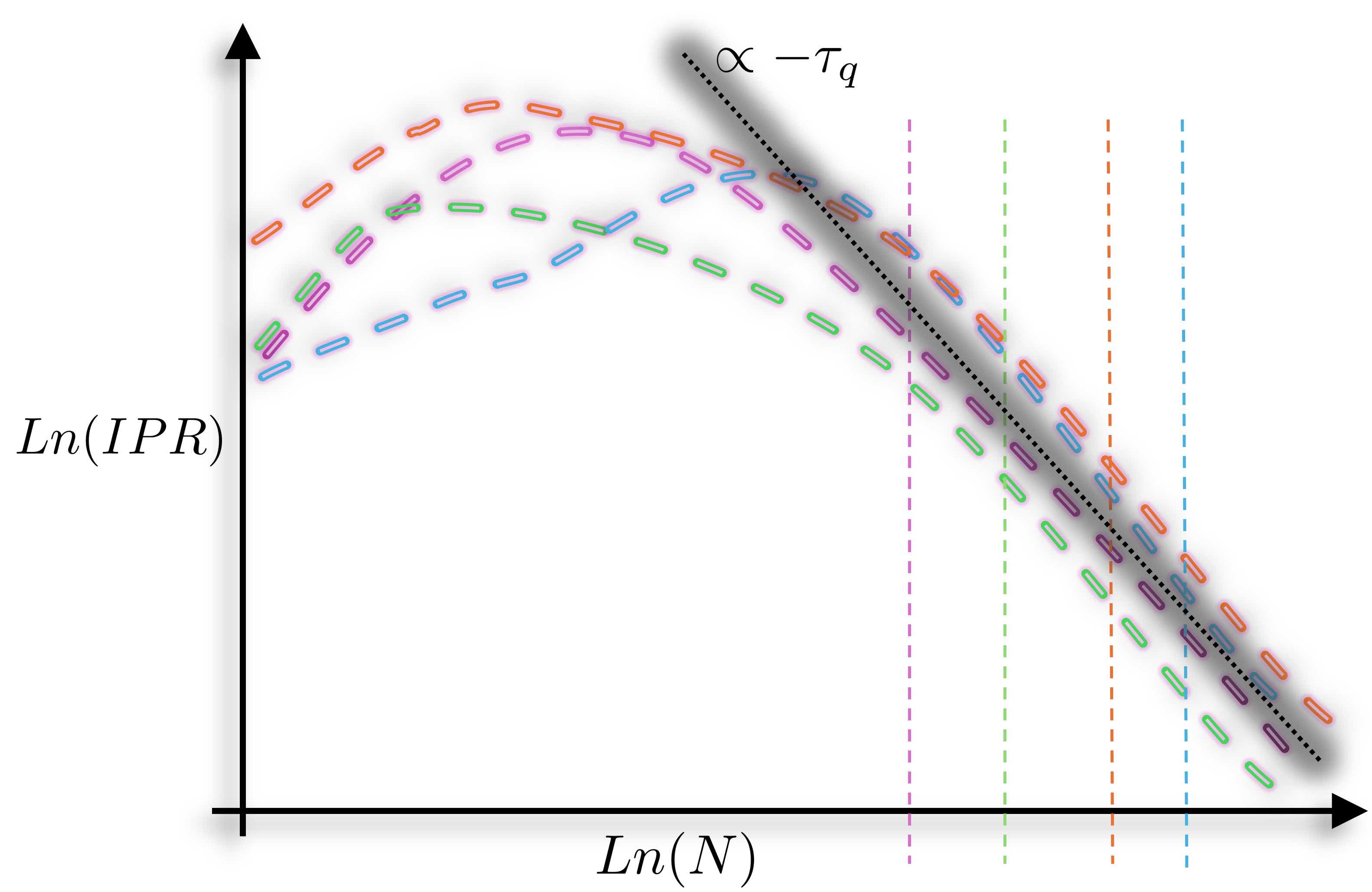}
    \caption{The overall behavior of the scaling of $\text{IPR}$ with the dimension of the Hilbert space of spin coherent states in the eigenbasis studied in this work. The dashed vertical colored lines indicate approximately the dimension at which a particular spin coherent state acquires its asymptotic behavior.}
    \label{fig:diagram}
\end{figure} 

The general behavior of the scaling of the $\text{IPR}$ can be seen in Fig.~\ref{fig:diagram}. The gray solid curve represent a spin coherent states which picks up an asymptotic value for very low values of the dimension (all of the models analyzed have a dimension $N=2J+1$, and in most cases we will be investigating the scaling from $J=50$ to $J=5000$). Once an asymptotic has been found, depending of their placement, all other spin coherent states will behave typically like the other four colors, 
with
the vertical dashed lines 
indicating 
the individual values of $Ln(N)$ where each one picks the same asymptotic values. In general, we will be studying two asymptotic values which correspond to regular and chaotic dynamics. 
How fast a spin coherent state 
reaches the asymptotic value 
strongly depends on its closeness to critical points.

\section{Classical correspondence and spin coherent states}
\label{sec:3}

The phase space of quantum collective spin models can be defined by employing Bloch, also called atomic, or spin coherent states~\cite{Perelomov1977}, where an effective Planck constant goes as $\hbar_{eff} = 1/J$. These are~\cite{coherent1}, 
\begin{equation}
    |z\rangle \equiv \dfrac{1}{(1+|z|^{2})^{J}}e^{z\hat{J}_{+}}|J,-J\rangle,
    \label{eqn:coherentstates}
\end{equation}
where $z = (Q+iP)/\sqrt{4-(Q^{2}+P^{2})}$. $\hat{J}_{+} = \hat{J_{x}} + i\hat{J_{y}}$ is the raising operator~\cite{esqptcollectivespin1}, 
$\hat{J}_{\alpha}$ are the components of the angular momentum vector, and they are generators of the group $\text{SU}(2)$ satisfying $[\hat{J}_{\alpha},\hat{J}_{\beta}] = i\varepsilon_{\alpha\beta\gamma}\hat{J}_{\gamma}$ ~\cite{Koczor_2019}. $J$ is the magnitude of the total angular momentum operator $\hat{J}^{2}$. The Hilbert space of the system is spanned by $2J+1$ basis vectors, so it is natural to employ the Dicke basis $|J,m\rangle$, where $\hat{J}^{2}|J,m\rangle = J(J+1)|J,m\rangle$ and $ \hat{J}_{z}|J,m\rangle = m|J,m\rangle$~\cite{haake1}. Here, we restrict the analysis to only integer values of $J$.

A corresponding classical model for a collective spin model can be obtained by taking the expectation value of the quantum Hamiltonian with respect to the coherent states in the thermodynamic limit $J\rightarrow\infty$ ~\cite{suncoherentstates1}
\begin{equation}
    H \equiv \lim_{J\to\infty} \dfrac{\langle z|\hat{H} |z \rangle}{J} \, .
  \label{eqn:classicalhamiltonian}
\end{equation}
A classical vector is defined in the same fashion $\vec{S} \equiv 
\langle\hat{J}_{z}\rangle/J = (S_{x},S_{y},S_{z})$. The coordinate transformations are $S_{x} = Q\sqrt{1-\frac{Q^2+P^2}{4}}$, $S_{y} = P\sqrt{1-\frac{Q^2+P^2}{4}}$ and $S_{z} = \frac{Q^2+P^2}{2} - 1$; which correspond to a stereographic projection of the Bloch sphere to the plane with the convention that the North pole corresponds to the center and the South pole to the circumference. Coherent states make contact with classical dynamics as $J$ increases and serve as a powerful tool to analyze specific regions of phase space by centering the coherent state at the point in phase space of interest~\cite{Bastarrachea2016,Wang2021,Wang2023,Bastarrachea2024}.

In the three following sections we explore the general properties of the scaling behavior of spin coherent states in three different bases. In the three systems, similar behavior of the $\text{IPR}$ of the coherent states is observed as the dimension $N$ is increased: an initial non-power law behavior, depending on the proximity of their coordinates with a critical point, followed in most cases by a similar power-law. The first model describes simple rotations around the $z$ axis, the bounded Harmonic Oscillator, whose eigenstates form the Dicke basis. The second is the Lipkin-Meshkov-Glick model, which has regular dynamics. The third is the QKT, which allows the exploration of the transition to chaos.

The analysis of the scaling in the Dicke basis, being simple, with only one critical point, helps to establish the methodology employed to study the following models. The LMG model has more critical points and has regular classical dynamics because the total angular momentum and the energy are conserved quantities, 
with a time-independent 
interaction. 
Next, the QKT model extends the LMG model, introducing a time-periodic function in the interaction term. In this case, the energy is not conserved, and the classical system is usually described in the literature as having 1.5 degrees of freedom \cite{Korsch1997}, exhibiting a transition from regular to chaotic behavior as a function of the kicking strength. In the three cases, the same overall qualitative behavior is found on the scaling of the $\text{IPR}$ of the coherent states, with the addition in the regions of chaotic motion of a different power-law exponent.
\section{Finite-size scaling for regular dynamics}
\label{sec:4}
\begin{figure}[t]
    \centering
      \includegraphics[width=0.95\columnwidth]{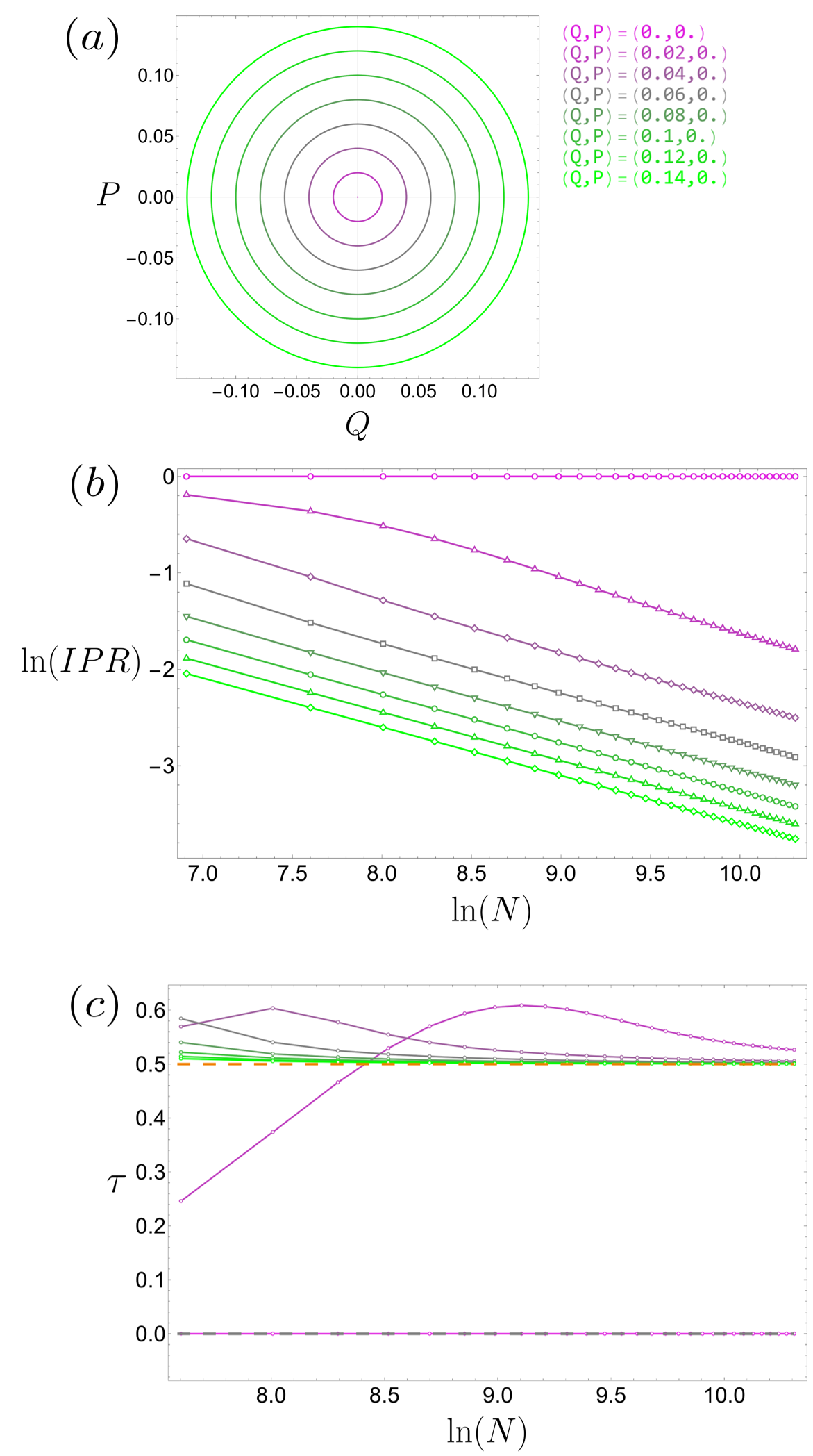}
    \caption{(a) Classical trajectories in phase-space of the eight initial conditions chosen as coordinates for the coherent states evolving under $H_{HO}$ (see text for details). (b) Log-Log plots of the $IPR$ of the same coherent states as a function of the dimension of the Hilbert space $N=2J+1$, $J$ varies from 500 to 15,000 in steps of 500. (c) Linear-log plots of the $\tau$ of the same coherent states as a function of the dimension (see text for details). The orange dashed line corresponds to the value of 0.5, and the bottom gray dashed line to the value of 0.}
    \label{fig:2}
\end{figure} 

We start analyzing the $\text{IPR}$ of coherent states in the Dicke basis $|J,m \rangle$, a natural representation given that coherent states are built as a rotation of the fiducial state that serves as the vacuum for creating Dicke states~\cite{coherent1}. Here and throughout this work, we will be considering the entire Hilbert space, even though the three Hamiltonians analyzed commute with the Parity operator $\hat{P} = e^{-i\pi\hat{J}_{z}}$. The reason for this choice is more obvious when considering coherent states, which will significantly overlap with the origin of coordinates that correspond to the Dicke state $|J,-J\rangle$. For integer values of $J$, there will be $J+1$($J$) Dicke states with positive (negative) parity, and the Dicke state $|J,m=0\rangle$ will always belong to the positive parity. This creates an unnecessary scaling factor to consider when analyzing other regions of phase space. The second reason is that it is known that a parity projected coherent state will be a superposition of two coherent states, as is exemplified in the LMG model, where the ground state is a superposition of two coherent states centered in each well, presented in the next section.

The Dicke basis spans the eigenvectors of the Hamiltonian $\hat{H}_{HO} = \alpha\hat{J}_{z}$, which describes rotations around the z-axis. Using the procedure described in \ref{sec:3}, its associated classical Hamiltonian is built, taking the expected value with respect to the coherent states. It is 
\begin{equation}
H_{HO}(Q,P)=\alpha\left(\frac{Q^{2}+P^{2}}{2}-1\right). \label{H0}
\end{equation}
It is the Hamiltonian of a (bounded) simple Harmonic oscillator in the canonical conjugate variables satisfying the Poisson bracket relation $\{Q,P\}=1$. There is only one critical point that corresponds to the ground state, a stable center, and no other relevant structure is present~\cite{esqptcollectivespin1}. Classically, if an initial condition is placed at the center, it will remain there, and all other initial conditions will revolve around this fixed point. 
In Fig.~\ref{fig:2} (a), the classical trajectories of eight initial conditions are depicted, with coordinates starting from $Q=0$ to $Q=0.14$ in increments of 0.02 for a representative value $\alpha=0.84$. 

To perform a similar analysis in the quantum regime, the behavior of $\ln(\text{IPR})$ vs. $\ln(N)$ is shown in Fig. \ref{fig:2} (b), for the same coordinates, now employed to define the coherent states, and with the same color code. The $\text{IPR}$ of the first coherent state corresponds to the first pink horizontal line at the top of Fig.~\ref{fig:2} (b). This is a consequence of the fact that, for any dimension, the coherent state with coordinates $(0,0)$ corresponds to the Dicke state $|z(0,0)\rangle=|J,-J\rangle$. This behavior is associated with the coherent state that does not scale; in other words, its power law has a null exponent. This will happen for any coherent state which only takes a constant amount of elements of the basis into consideration, as a function of $N$. The second curve corresponds to the coordinate $(Q,P)=(0.02,0)$. It can be seen that, from $Ln(N)=7$ to $Ln(N)=9$, the $\text{IPR}$ exhibits a curved, a non-power law behavior, and transitions to a linear, power-law one, for larger values of $N$. As this coherent state $|z(0.02,0)\rangle$ is described by a small fraction of elements of the Dicke basis, it requires a large dimension to express the power law behavior. The closer a coherent state is to the center, the more non-linear behavior the $\text{IPR}$ curve will show in the log-log plot. The remaining coherent states very quickly pick up the same power-law behavior as a function of $N$.

To get a more quantitative description of the previous analysis, in Fig.~\ref{fig:2} (c), we show the finite mass exponent $\tau$ of the eight coherent states as a function of the dimension. In Fig.~\ref{fig:2} (b), we connected each point with linear segments. The slopes of each segment are the values of the finite mass exponent. When the $\text{IPR}$ shows a power-law behavior in the log-log plot, $\tau$ is the exponent of this 
power-law. If this is not the case, then the finite mass exponent only represents the change in the proportion of the Dicke states that the coherent state is using when the dimension is increased. As mentioned in the previous paragraph, the power-law of the first coherent state is exactly zero, and its $\tau$ curve is located at the bottom. From top to bottom, the second curve corresponds to the coherent state $|z(0.02,0)\rangle$. It is the last one to start picking up the same power-law of the other six coherent states: for $\ln{N}=\ln{30001}=10.3$ the value of $\tau$ has almost reached the limiting value of 0.5.
\begin{figure}
    \centering
    \includegraphics[width=\columnwidth]{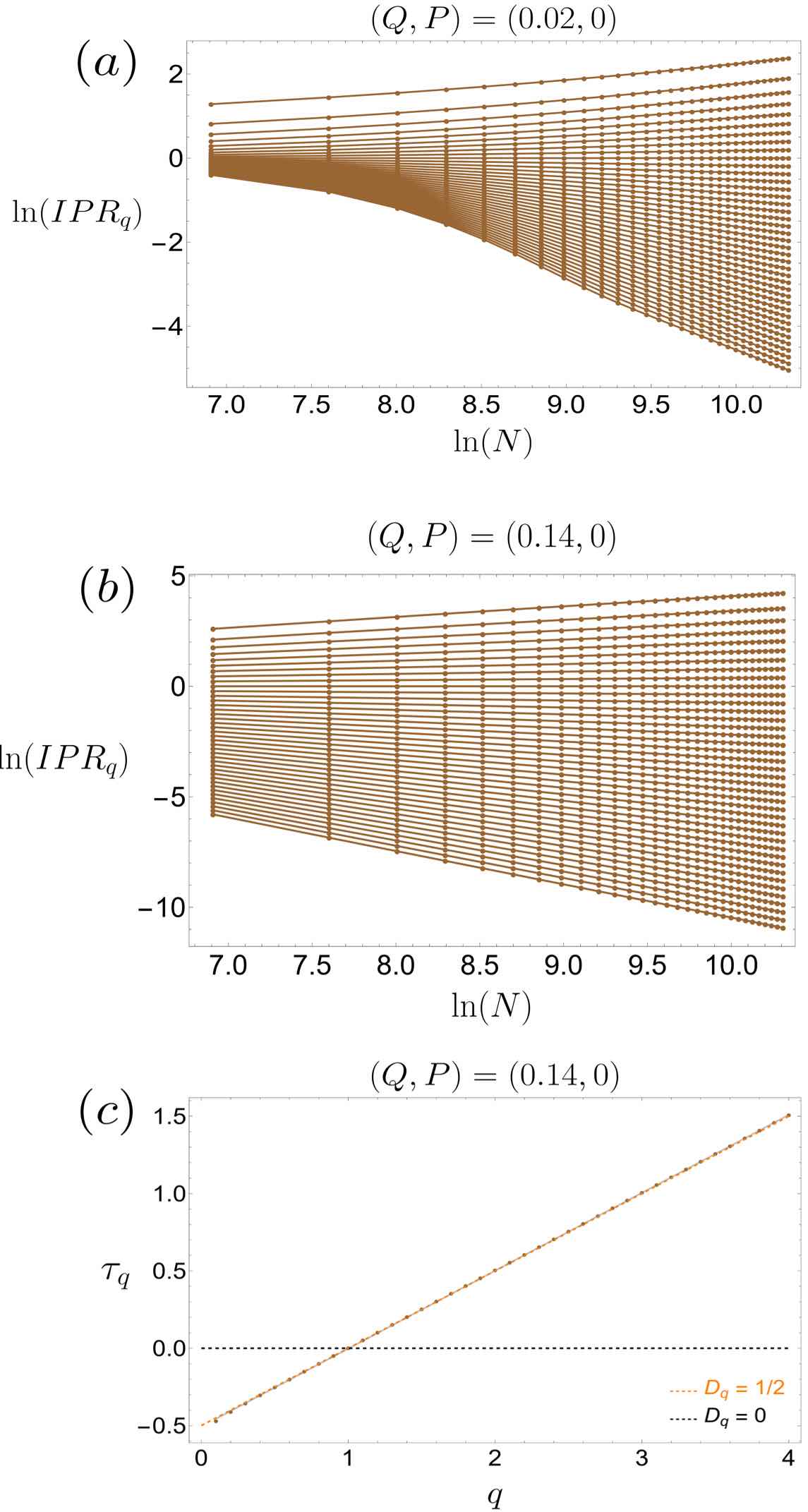}
    \caption{(a) Log-log curve of the $\ln{\text{IPR}_{q}}$ vs $\ln{\left(N\right)}$ of the coherent state $|z(0.02,0)\rangle$. (b) Log-log curve of the $\ln{\text{IPR}_{q}}$ vs $\ln{\left(N\right)}$ of the coherent state $|z(0.14,0)\rangle$. In (a) and (b), the values of $q$ go from $q=0.1$ to $q=4.0$ in steps of 0.1 from top to bottom. (c) $\tau_{q}$ vs $q$ of the coherent state $|z(0.14,0)\rangle$. The orange dashed line corresponds to a mono-fractal state, which scales as a Gaussian distribution~\cite{Bastarrachea2024}, and the black dashed horizontal line corresponds to a quantum state, which does not scale with the dimension $N$.}
    \label{fig:3}
\end{figure} 

Up to this point, we have only studied the power-law behavior captured by the finite mass exponent for the case of $\text{IPR}_{q=2}$. A multifractal characterization requires a similar analysis for other parameter values $q$. We explore this in the interval for $q \in [0.1,4]$. Employing values of $q$ closer to zero or higher than 4 requires expanding the numerical precision and does not provide relevant additional information in the present context. In Fig.~\ref{fig:3}(b) we show the $\text{IPR}_{q}$ for the last initial condition, with coordinates $(Q,P)=(0.14,0)$. When $\text{IPR}_{q}$ exhibits a power law behavior, its slope is proportional to $q-1$, as shown in Eq. \ref{eqn:Dq}. The curves with 
negative
slope correspond to the values with $q<1$, and the curves with 
positive
slope to the values with $q>1$; the last two kinds of slopes are separated by the horizontal line with zero slope, $q=1$, which is the normalization condition. The $\text{IPR}_{q}$ of the second coherent state is shown  in Fig.~\ref{fig:3} (a). In the range of dimensions employed, from $J=500, N=1001$ to $J=15000, N=30001$, for all values of $q$ there is no linear relation between $Ln(\text{IPR}_{q})$ and $Ln(N)$, no power law behavior. It implies that it is not possible to perform a formal MFA in this state. From observing the other coherent states, we conclude that the closer the coherent state is to the critical point, the greater the dimensions needed to perform the MFA.

To summarize, in Fig.~\ref{fig:3} (c) the slopes $\tau_q$ of all the lines in Fig.~\ref{fig:3} (b) are plotted against $q$. The linear relation with $q$ is clearly visible, allowing us to employ Eq. \ref{eqn:Dq} to derive the fractal dimension $D_0 = \frac{1}{2}$. It is a clear example of monofractal scaling. This line is representative of nearly all other initial conditions. The exceptions are the first one, which does not scale with the dimension and is equivalent to the horizontal black dashed line, and the second one, where the MFA cannot be performed for the range of dimensions explored in this work. The most significant learning from this section is the usefulness of the finite mass exponent $\tau$. It indicates when the dimension is large enough to perform a multifractal analysis (MFA): in the region where a power-law scaling of the $\text{IPR}_{q}$ is observed for all the values of $q$ of interest. In the case of the $\text{IPR}$ of the coherent states in the Dicke basis, the second coherent state $|z(0.02,0)\rangle$ does not exhibit a power-law in the range of dimensions from $N=1001$ to $N=30001$. It has been shown analytically that a coherent state with a Gaussian distribution in a particular basis, associated with regular classical dynamics, in the limit of $N\gg1$, has a $D_{0}=1/2$~\cite{Bastarrachea2024}. In numerical studies, the challenge is to employ a dimension large enough to exhibit this power-law scaling for a given coherent state. The finite mass exponent helps determine when a formal MFA analysis can be performed in each case.

It is worth mentioning that there is more information on the dependence of the finite mass exponent with $N$ than that MFA can capture. When it is not possible to perform an MFA, the absence of a power-law provides valuable information about the location of the coherent state in phase space. We show elsewhere~\cite{Miguel2025} how the dynamics of a coherent state located close to a fixed point leaves a fingerprint in the scaling of the \text{IPR}, even in the absence of a power-law scaling and without the need to go to the limit of $N\gg1$.

\section{Finite size scaling in the LMG model}
\label{sec:5}

\begin{figure}[t]
    \centering
    \includegraphics[width=\columnwidth]{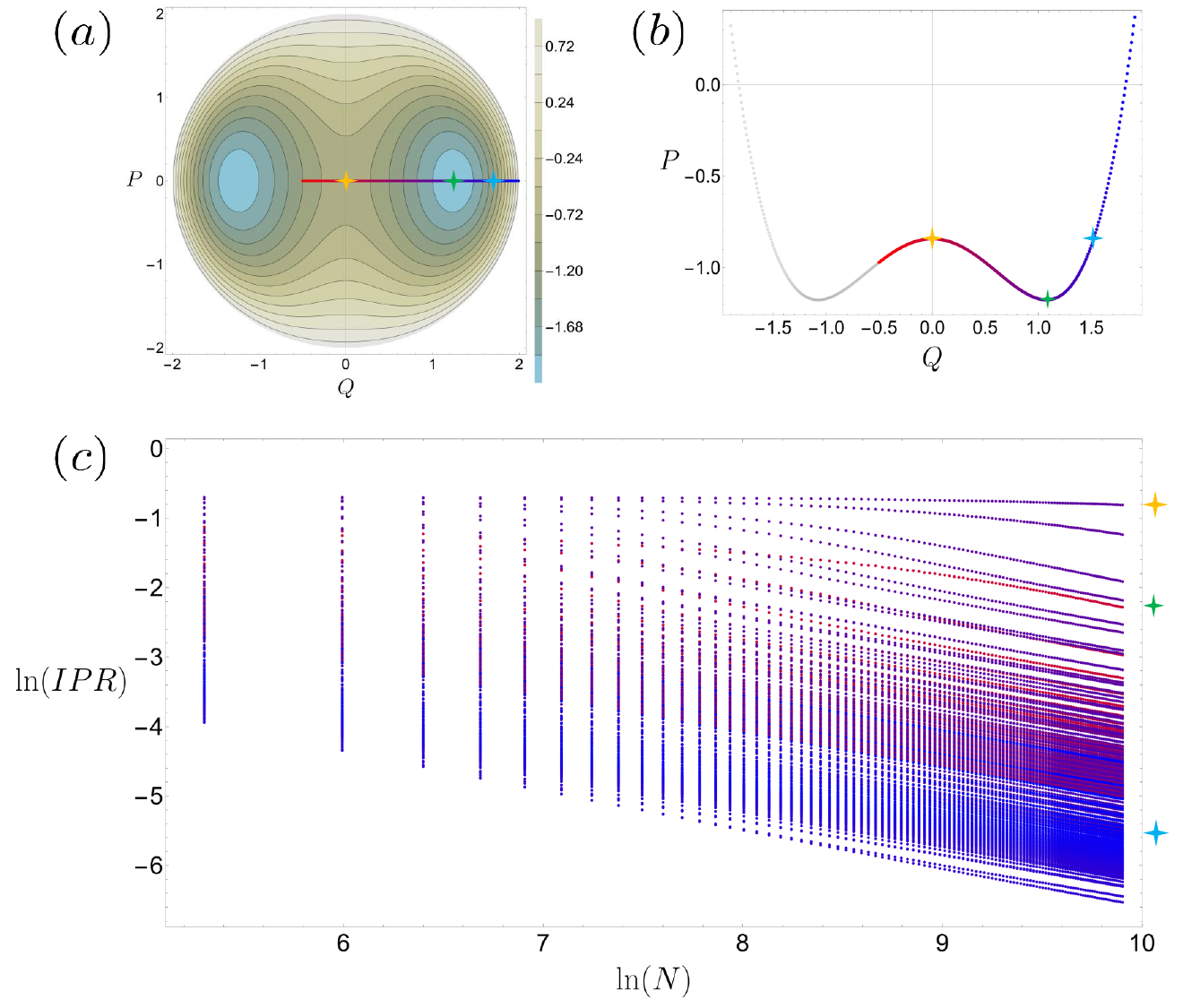}
    \caption{(a) Contour plot of the semiclassical energy function of the LMG model for $\alpha=0.84$ and $k=-2$. (b) Classical energy cut from Fig.~\ref{fig:lmg} (a) at $P=0$. The coordinates of 250 coherent states are shown as each colored dot. The yellow cross indicates the unstable fixed point, the green one indicates the stable minimum, and the cyan indicates the separatrix edge where there is an ESQPT. The gray dots are shown only for reference. (c) Log-Log plot of the $\text{IPR}$ as a function of $N$ for the 250 coherent states in the LMG basis. The dimension goes from $N=201$ to $N=11001$ in increments of $J=100$.}
    \label{fig:lmg}
\end{figure} 

In what follows, we present a detailed study of the scaling of the $\text{IPR}$ for selected coherent states $|z(Q,P)\rangle$ in Eq.~\ref{eqn:coherentstates}, representing points in the phase space associated with the dynamics of the Lipkin-Meshkov-Glick Hamiltonian~\cite{Lipkin1965,Meshkov1965,Glick1965} (LMG), a model from nuclear physics and quantum magnetism that have become a customary tool to study collective qubit systems in several fields including such as cold atoms~\cite{Chen2009,Sauerwein2023,Muniz2020} or cavity QED setups~\cite{Morrison2008}, whose critical properties and classical limit have been extensively explored~\cite{Dusuel2004,Castanos2006,Heiss2006,Ribeiro2007,Ribeiro2008}. The LMG model corresponds to the non-kicked limit of the QKT, which will be analyzed later. It describes the continuous motion of an angular momentum vector with rotation and torsion. Its Hamiltonian is ($\hbar=1$):
\begin{equation*}
\hat{H}_{LMG} = \alpha\hat{J}_{z} + \dfrac{k}{2J}\hat{J}_{x}^{2} .
\end{equation*}
Its semiclassical limit, obtained via spin coherent states, reads
\begin{equation}
H_{LMG} = \alpha S_{z} + \frac{k}{2}S_{x}^{2}.
\end{equation}

In what follows, we set the values $\alpha=0.84$ and $k=-2$. The classical dynamics is regular, and for the selected parameters, three critical points emerge: two stable minima and one unstable saddle point with a positive Lyapunov exponent ~\cite{positivelyapunov1}. In Fig.~\ref{fig:lmg} (a), the corresponding classical energy surfaces of the LMG model are shown, consisting of a double well and a local maximum. Fixed points are crucial in our analysis because the scaling we investigate gets us closer to the semiclassical limit. The coherent states provide a mean-field description of these critical points, which becomes exact in the thermodynamic limit $N \rightarrow \infty$~\cite{suncoherentstates1, phasespacemethod1, esqptcollectivespin1, phasespacegeometry1}. 
\begin{figure*}[t]
    \centering
    \includegraphics[width=\textwidth,height=6.4cm]{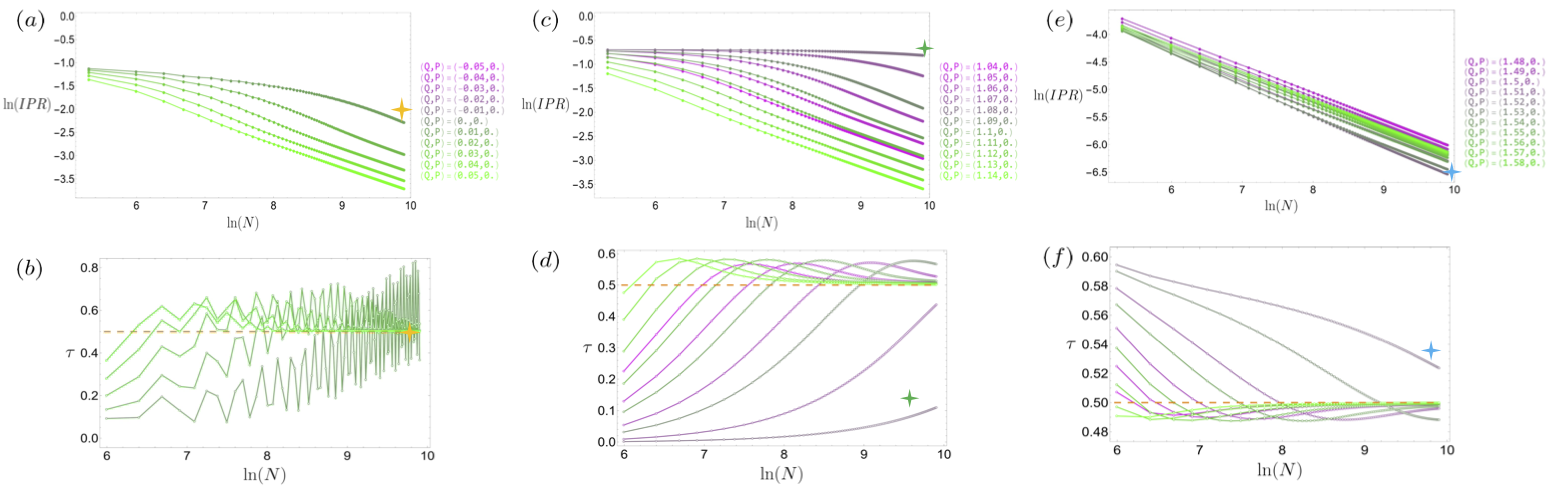}
    \caption{Finite Size Scaling in the LMG Model. (a) Log-Log curves of the $\text{IPR}$ vs $N$ eleven coherent states with coordinates $P=0$ and $Q$ from 1.04 to 1.14 in steps of 0.01. (b) (linear-linear) Finite mass exponent as a function of $\ln{\left(N\right)}$ for the same eleven coherent states. The coherent state closest to the stable fixed point (right minimum in Fig.~\ref{fig:lmg}) is marked with the green cross. (c) Same as (a) but for eleven coherent states with coordinates $P=0$ and $Q$ from 1.48 to 1.58 in steps of 0.01. (d) Same in (b) but for the same eleven coherent states in (c). The coherent state closest to the separatrix edge is marked with the cyan cross. (e) Same as (a) but for ten coherent states with coordinates $P=0$ and $Q$ from -0.05 to 0.05 in steps of 0.01. (f) Same as (b) but for the same eleven coherent states in (e). The coherent states next to the unstable fixed point are marked with the yellow cross $|z(\pm0.01,0)\rangle$. Due to the symmetry of the location of the coherent states, the pink curves in both panels are exactly under the green ones. The diagonalization was done from $\ln{\left(N\right)} = \ln{\left(2(100)+1\right)}$ = 5.30 to $\ln{\left(N\right)} = \log{2(10000)+1}$ = 9.90 in steps of $J=100$. The dashed horizontal orange line in (b), (d), and (e) corresponds to the value of 0.5.}
    \label{fig:5}
\end{figure*} 
Fig.~\ref{fig:lmg} (b) displays a cut of the energy surface $H_{LMG}$ for $P=0$. Each dot represents the coordinates of one of 250 coherent states selected along a segment from $(Q,P)=(-0.5, 0)$ to nearly the edge of the phase space $(Q,P) = (1.99, 0)$. In Fig.~\ref{fig:lmg} (c), we plot $\ln(\text{IPR})$ versus $\ln(N)$ with the same color code for each coherent state in Fig.~\ref{fig:lmg} (b). By looking at their asymptotic behavior, 
most of these coherent states show a clear linear behavior, with their $\text{IPR}$ scaling with the power-law $N^{-1/2}$, with a finite mass exponent $\tau= 0.5$. As mentioned above, this slope has been associated before with a coherent state that scales as a Gaussian when distributed over the eigenstates of a Hamiltonian~\cite{Bastarrachea2024}. While it is not guaranteed that the components of an atomic coherent state will have a Gaussian profile when expressed on the LMG basis, we have confirmed that this is the case for all the particular coherent states we investigated. 

There is a set of coherent states whose values do not change as fast as a function of $N$. Their curves are those at the top in Fig.~\ref{fig:lmg} (c). This set encloses those coherent states centered closer to the fixed points and the separatrix, denoted by the three different crosses in Fig.~\ref{fig:lmg} (b)]. They seem to require a greater dimension to display the same power-law behavior of $N^{-1/2}$. In these cases, the asymptotic limit has not been reached for these values of $N$, which are very large, so there is not a distinguishable power-law behavior of the $\text{IPR}$, a mass exponent $\tau_{q}$ cannot be obtained, making not possible to perform a formal multifractal analysis in this range of dimensions. The first curve corresponds to the coherent state, which is closest to the right minimum in the energy surface [Fig.~\ref{fig:lmg} (a)] indicated by the yellow cross. For large $N$, it becomes very close to the exact ground state. Notice that the exact quantum ground state combines the right and left coherent states at the minima due to symmetry. This state has $\text{IPR}\sim 2$ and remains essentially constant, with a null slope, scaling as $N^{0}$. Inspired by Ref.~\cite{apparentdimension1}, we perform a detailed scaling analysis around the two critical points and the separatrix edge to better understand these coherent states.

In Fig.~\ref{fig:5} (a), we show the log-log curves of the $\text{IPR}$ vs. $\ln{N}$ and in panel (b), the finite mass exponent of eleven coherent states around the stable fixed point. Most of these coherent states overlap significantly with many eigenvectors with energies close to the minimum, except for the one associated with a single eigenvector, which corresponds to the ground state. The $\text{IPR}$ curves are very smooth as a function of the dimension. They have qualitatively the same behavior as those found in the Dicke basis. The coherent state located near the right minimum $|z(1.08,0)\rangle$ is the state that has not reached its asymptotic power law at the dimension considered here $N=20001$. We believe that if the coherent state is not located exactly at the fixed point for sufficiently high dimensions, it will acquire the same power-law of $\text{IPR}\approx N^{-0.5}$.

In figure \ref{fig:5} (c) and (d), we analyze another eleven coherent states in detail, but now focusing on the region of the separatrix edge, where there is an Excited State Quantum Phase Transition (ESQPT) in the LMG model~\cite{Cejnar2021,Engelhardt2015}. In this case, we get similar results as in previous sections: some coherent states have reached their asymptotic values with slope $1/2$, while others would require larger dimensions to arrive at this value. A detailed study of the connection of the ESQPT and the specific behavior of the scaling of the $\text{IPR}$ is beyond the scope of the present article.

In Fig.~\ref{fig:5} (e) and (f), we show the scaling of the $\text{IPR}$ and its corresponding finite mass exponent $\tau$ of five coherent states to the left of the unstable fixed point, located at $(Q,P)=(0,0)$, and five to the right. Due to the symmetry of their location, only the green curves can be seen. In this case, we observe exactly the same qualitative behavior of the $\text{IPR}$ of the coherent states located around a critical point as in the previous section. The closer the coherent state is to a critical point, the closest being these two: $|z(\pm0.01,0)\rangle$, the larger the dimension needed to exhibit the power-law $\text{IPR}\approx N^{-0.5}$.
The slope between two consecutive points in Fig. \ref{fig:5} (e) is the finite mass exponent, whose dependence on $Ln(N)$ is shown in Fig.~\ref{fig:5} (f). It exhibits strong fluctuations around the limiting value of $\tau=0.5$. These sudden changes are associated with the fact that any coherent state with coordinates close to $(Q,P)\approx(0,0)$ has very few elements in the Dicke basis, and also in the $\hat{H}_{LMG}$ basis, which makes the $\text{IPR}$ to have significant changes as $N$ is increased. Although in the thermodynamic limit $N\to \infty$, there will be only one eigenvector at the critical point energy, for any finite value of $N$, the exact location and energy of this eigenvector usually fluctuate around the energy of the coherent state $|z(0,0)\rangle$. This is a crucial difference between the critical points analyzed in the previous section because the center of the phase space in the LMG model is now an unstable fixed point. 
A detailed analysis relating the fluctuations of the $\text{IPR}$ quantified by $\tau$ and the stable or unstable fixed points will be investigated elsewhere~\cite{Miguel2025}.

\begin{figure}[h]
    \centering
    \includegraphics[width=\columnwidth,height=6cm]{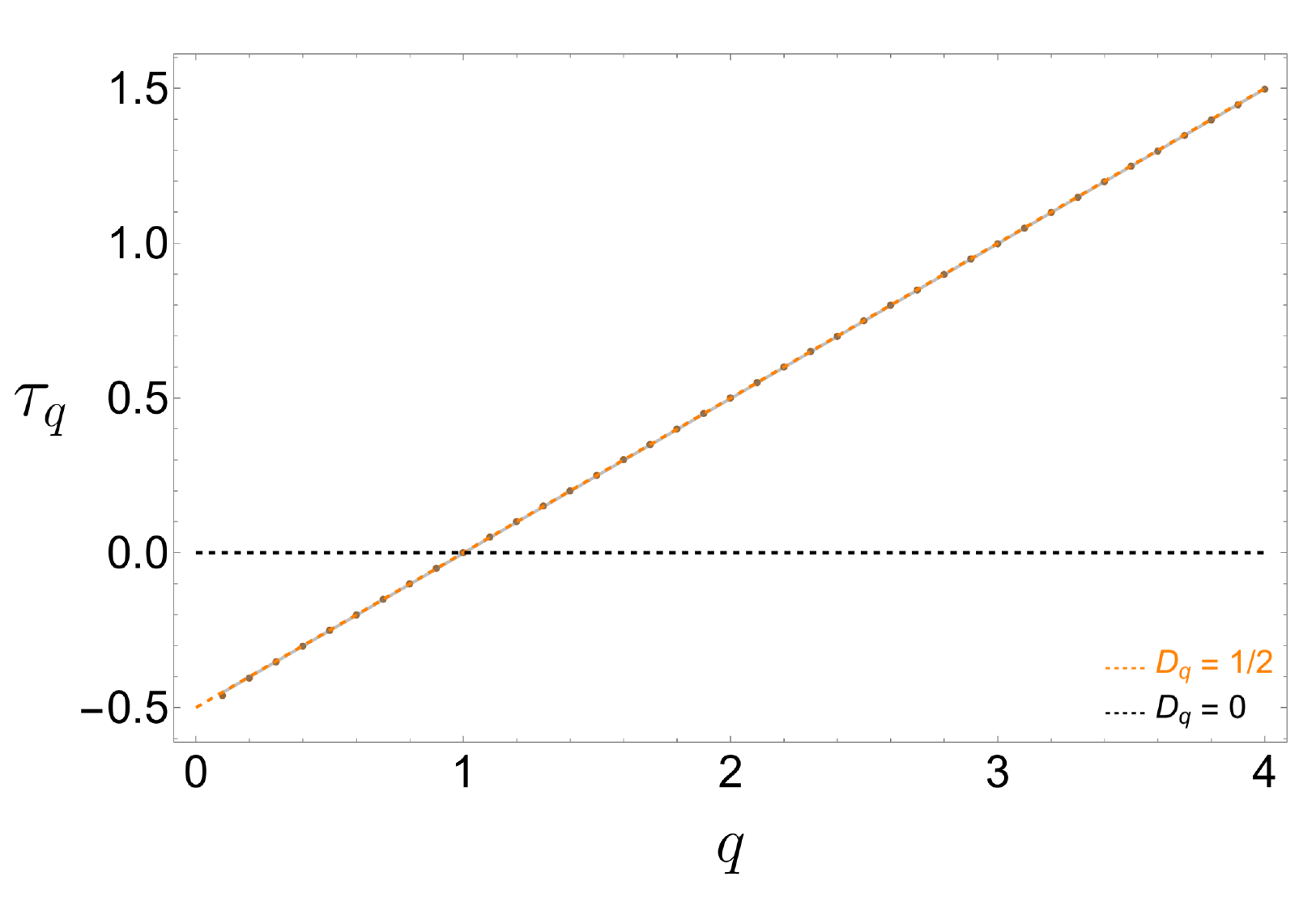}
    \caption{$\tau_{q}$ vs $q$ of the coherent state $|z(-0.5,0)\rangle$
    for the LMG model. The orange dashed line corresponds to a mono-fractal state, and the black dashed horizontal line corresponds to a quantum state that does not scale.}
    \label{fig:taulmg}
\end{figure}

Finally, Fig.~\ref{fig:taulmg} shows the MFA analysis for one representative coherent state expressed in the LMG basis. It gives the same information as the Dicke basis: the coherent state is located in a regular region. All other initial conditions analyzed in which a MFA can be performed with the conditions mentioned before give exactly this same figure.

\section{Scaling in the Quantum Kicked Top}
\label{sec:6} 

The quantum kicked-top model is one of the simplest quantum systems that allows the study of the transition from regular to chaotic dynamics as the kicking strength increases ~\cite{haake1,haakesymmetries2,haake3,Santhanam2022}. Its advantage lies in its algebraic simplicity, finite dimension, and well-defined semiclassical limit. The quantum kicked top model has been experimentally realized using the combined electronic and nuclear spin of a single atom\cite{Chaudhury2009}, where deep in the quantum regime, a good correspondence between the quantum dynamics and classical phase space structures was observed.  The dynamics of a mean-field quantum kicked top can be simulated by performing weak collective measurement on a large ensemble of two-level quantum systems and applying global rotations conditioned on the measurement outcome \cite{Munoz2020}. The presence of chaos can be observed by measuring the out-of-time-order correlation functions \cite{Swingle2016,Rozenbaum2017}. Effective thermalization \cite{Altland2012}, and ergodic dynamics have been observed in a small quantum system consisting of only three superconducting qubits \cite{Neill2016,Madhok2018}. The fixed-point analysis and calculation of the Lyapunov exponents show correspondence to the random matrix theory in the chaotic regime \cite{Ray2016}. 

The Hamiltonian of the QKT includes a $z$-axis rotation of a quantum angular momentum vector $\hat{\vec{J}}$ and kicked torsions around the $x$-axis, given by ($\hbar$=1) ~\cite{haake1}
\begin{equation}
    \hat{H}(\alpha,k,t) = \alpha\hat{J}_{z} + \frac{k}{2J}\hat{J}_{x}^{2}\sum_{n=0}^{\infty}\delta(t-n\tau) ,
    \label{eqn:QKT}
\end{equation}
The $z$-rotation is parametrized by $\alpha$, and periodical torsions are modeled by a train of delta functions with strength controlled by $\frac{k}{2J}$.

\subsection{Floquet formalism}
\label{sec:floquet} 

Because the Hamiltonian is periodic in time, one can apply the Floquet theorem~\cite{Sambe1} to calculate a state's temporal evolution at times $t_{n}=n$, considering a unitary time scale $\tau=1$. The time-evolved state is given by $|\psi(t_{n})\rangle = \hat{F}^{n}|\psi(0)\rangle \,$, where the Floquet operator $\hat{F}$ is
\begin{equation}
    \hat{F}(\alpha,k) = \exp\left[-i\alpha\hat{J}_{z}\right]\exp\left[\frac{-ik}{2J}\hat{J}_{x}^{2}\right] \, .
    \label{eqn:floquet}
\end{equation}
The Floquet eigenvectors $|\phi_{j}\rangle \,$
fulfill the equation $ \hat{F}|\phi_{j}\rangle = e^{i\phi_{j}}|\phi_{j}\rangle \,$, where the eigenvalues $\phi_{j}$ (mod $2\pi$) are called {\it quasienergies}~\cite{haake1}. Due to the unitary character of the Floquet operator, quasienergies are complex, with unitary modulus, and distributed on the unitary circle in the complex plane. Depending on the chosen branch, they can be listed in any interval of size $2\pi$, like $[-\pi,\pi]$ or $[0,2\pi]$. Because the Floquet operator commutes with the parity operator $\hat{P} = e^{-i\pi\hat{J}_{z}}$, $[\hat{F},\hat{P}] = 0$, it does not change the parity of well-defined parity initial state. For instance, the Dicke basis and the eigenvectors of $\hat{F}$ have well-defined parity, while the atomic coherent states do not. The model has several geometric symmetries, classical and quantum, affecting its dynamics~\cite{haakesymmetries2, haake4}. In Ref.~\cite{pspinmodels1}, the authors describe most of the classical symmetries of the QKT and related p-spin models. The overall impact of these symmetries depends heavily on the particular phenomena one is investigating. Also, depending on the parameters chosen, one may encounter other symmetries as the ones described in ~\cite{recurrences1}. When chaos is fully developed in the whole phase space, the quasienergy spectrum of the Floquet operator is very well characterized by the Circular Orthogonal ensemble (COE), as described in ~\cite{haakesymmetries2}, due to its anti-unitary time reversal invariance.

\subsection{Classical correspondence in the QKT}

In the classical limit, the system becomes a classical top fixed by its tip, free to rotate in the $z$ direction and experiencing periodical kicks along the $x$ axis~\cite{haake1, Wang2021}.
\begin{equation}
    H  = \alpha S_{z} + \dfrac{k}{2}S_{x} ^{2}\sum_{n=-\infty}^{+\infty}\delta(t-n\tau) \, ,
    \label{eqn:CKT}
\end{equation}
After each kick, the dynamics on the Poincaré section are given by the Poincaré map, which results from solving the classical equation of motion before and after the kicks~\cite{classicaltop1}. One then gets from the classical Hamiltonian in Eq.~\ref{eqn:CKT}, a nonlinear area-preserving map, the Poincaré map  $\vec{S}_{n+1} = \, F \vec{S}_{n}$ with $F$ given by
\begin{equation}
    F = \begin{pmatrix}
                \cos{(\alpha)} & -\sin{(\alpha)}\cos{(kS_{x})} & \sin{(\alpha)}\sin{(kS_{x})} \\
                \sin{(\alpha)} & \cos{(\alpha)}\cos{(kS_{x})} & -\cos{(\alpha)}\sin{(kS_{x})} \\
                0 & \sin{(kS_{x})} & \cos{(kS_{x})}
        \end{pmatrix}
        \label{eqn:floquetclasico}
\end{equation}

\subsection{Quantum chaos in the QKT}

\begin{figure}[H]
    \centering
    \includegraphics[width=0.95\columnwidth,height=5.25cm]{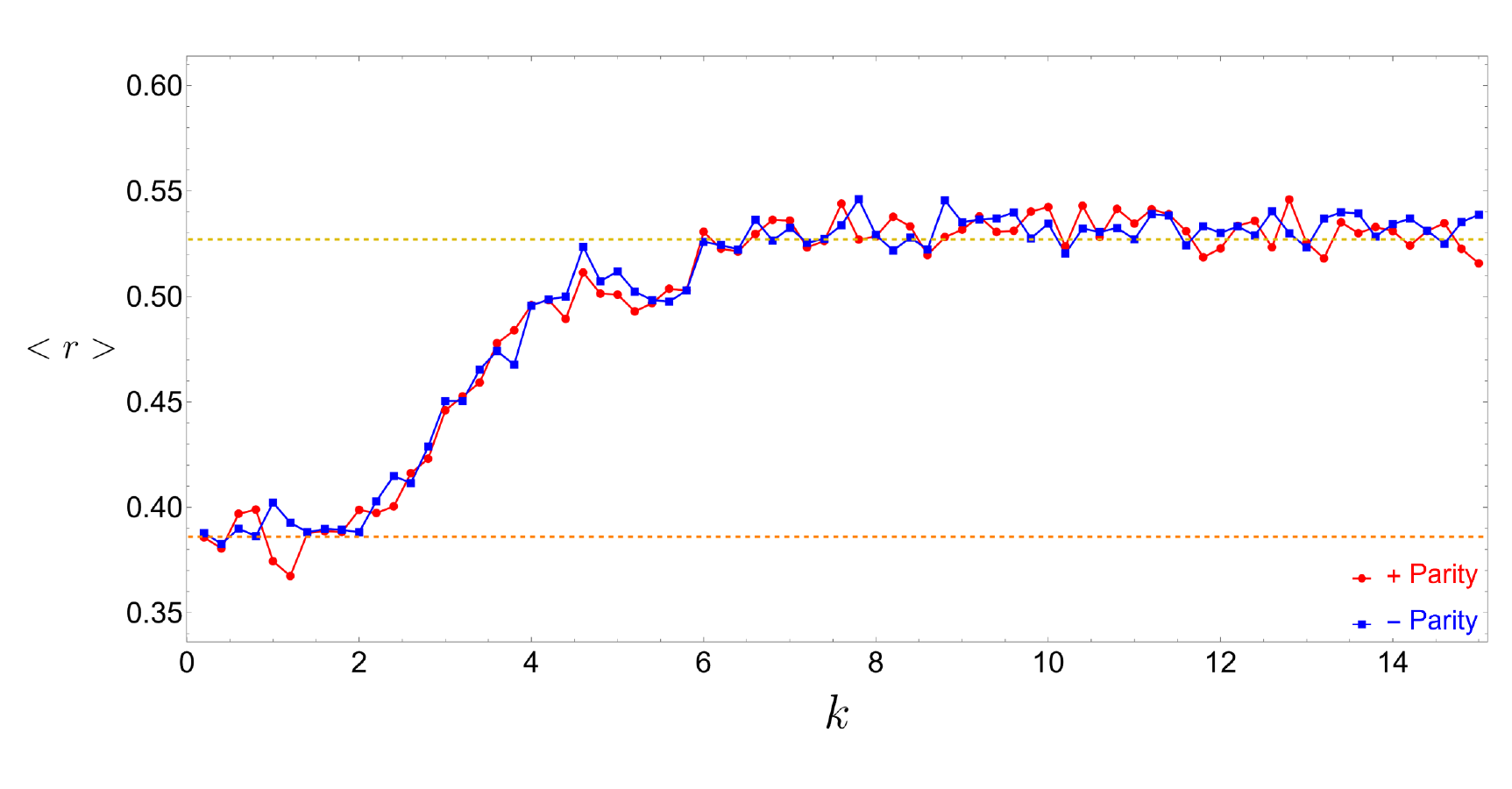}
    \caption{Mean value of $r$ as a function of $k$ for both parity sectors in the QKT. The dimension of the Floquet operator is $N = 4001$, and each parity sector is of dimension $N_{+} = 2001$ and $N_{-}=2000$. The orange dashed horizontal line marks Poissonian statistics, and the golden dashed horizontal line marks Wigner-Dyson statistics.}
    \label{fig:meanr}
\end{figure}

The quantum model is periodic in the kicking strength for integer and semi-integer $J$ integer~\cite{recurrences1}. Still, in either case, when chaos is fully developed in the phase space, both parity sectors of the Floquet operator used in this work always belong to the COE ensemble ~\cite{haakesymmetries2}. Throughout this work, we keep fixed the value $\alpha = 0.84$ and study the transition from regularity to chaos as a function of $k$. To quantify quantum chaos, a useful signature is the average $\langle r \rangle$ of the spectral parameter $r_{i} = \text{min}(1/\delta_{I},\delta_{I})$ with $\delta_{i} = (\phi_{i+1}-\phi_{i})/(\phi_{i}-\phi_{i-1})$~\cite{Wang2021}. The $r$ factor is calculated over the full quasienergy spectrum $\{ \phi_{i} \}$, separated by parity and sorted in ascending order. In Fig.~\ref{fig:meanr}, the value of $\langle r \rangle$ is shown as a function of the kicking strength $k$, for a fixed value of $\alpha$ and $J=2000$. The orange dashed line in Fig.~\ref{fig:meanr} at $r=0.386$ indicates the value associated with Poisson statistics, hence regular dynamics, and the golden dashed line at 0.53 corresponds to a Wigner-Dyson statistics of the spacings, associated with chaotic dynamics. It can be seen that the QKT  has essentially regular dynamics for $k<2$ and chaotic dynamics for $k>7$. Notice that there are subtle differences between each parity sector.

\begin{figure*}[t]
    \centering
    \includegraphics[width=\textwidth]{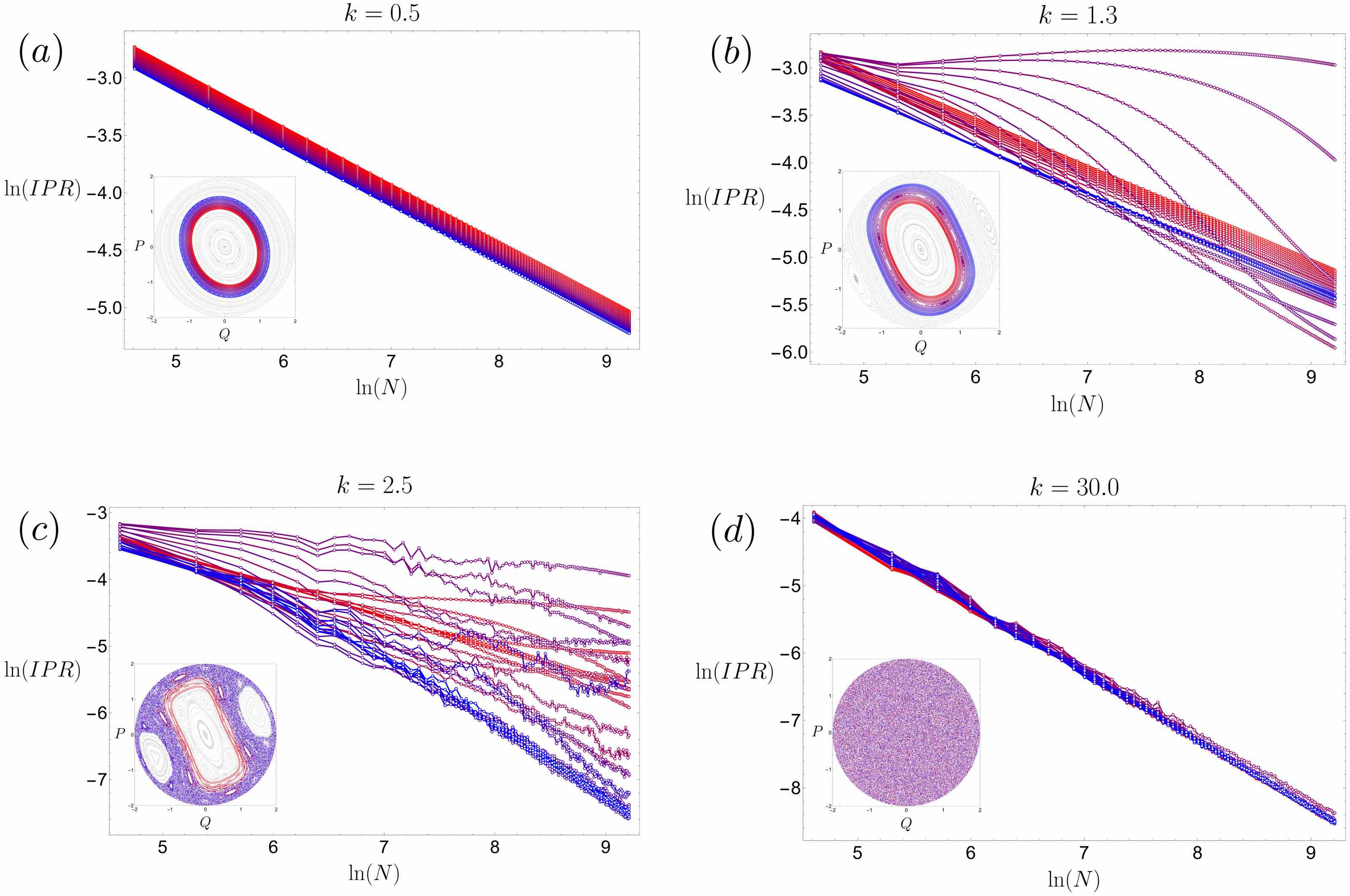}
    \caption{Log-Log plots of the $IPR$ vs $N$ of 31 coherent states in the QKT for four values of the kicking strength: (a) $k=0.5$, (b) $k=1.3$, (c) $k=2.5$, and $k=30$. For all initial conditions, $P=1$ and $Q$ vary from 0.2 to 0.8 in steps of 0.02. The insets show the classical Poincar\'e section, with each initial condition sharing the color with the IPR of the corresponding coherent state.}
    \label{fig:collageipr}
\end{figure*} 

\subsection{Finite size scaling in the QKT model}

There is a fundamental difference between the QKT and the LMG and rotor models. The description of the QKT model is stroboscopic in the classical and quantum regimes. So, the quantum procedure expresses the coherent states on the Floquet basis, not on a Hamiltonian basis. On the other hand, it has been shown in the literature that there is indeed a quantum-classical correspondence between the eigenvectors of the time-periodic description of the QKT model with the Floquet operator and the predictions of Random Matrix Theory in the chaotic regime, with the corresponding classical structures generated by the Poincaré map in the classical regime ~\cite{pusc1,MKus_1988,structuresqkt1,Wang2023}. 

While some results are similar to those of the previous sections, the scaling analysis presents additional challenges. At variance from the previous systems, whose dynamics are regular and have very few critical points, in the QKT model, there are many critical points. They are described by the Poincaré-Birkhoff theorem: when a resonant Tori breaks, it leaves behind a new set of fixed points ~\cite{chaosinmaps1, resonances1}), which are the classical structures surviving in phase space after a system is brought from regular to chaotic dynamics~\cite{structuresqkt1}. When the kicking strength $k$ is non-null, there are infinitely many critical points, and the scaling analysis could become cumbersome because, as it was shown in the previous sections, the proximity of critical points requires higher dimensions to rich the linear power law behavior where the MFA can be performed. When it is possible, in what follows, it is shown that, in the chaotic regime and for very large dimensions, there is a new exponent for the power-law scaling of the $\text{IPR}$, which captures the chaotic motion. In Fig.~\ref{fig:collageipr} it is shown the scaling of the $\text{IPR}$ of thirty-one coherent states for four values of the kicking strength, calculated from $J=50$ to $J=5000$ in steps of 50. The insets present the associated Poincar\'e sections.

\subsection{Generic regular power-law behavior of the $IPR$}

Because the $k=0.5$ and $k=1.3$ cases exhibit dominant regular dynamics, we anticipate regions where one obtains scaling results similar to those of the HO and LMG Hamiltonians. In Fig.~\ref{fig:collageipr} (a) and (b) it is shown the $\text{IPR}$ curves for thirty-one coherent states for two different values of the kicking strength $k=0.5$, and $k=1.3$. Most curves have already reached the same asymptotic power-law of $\text{IPR}\approx N^{0.5}$. In panel Fig.~\ref{fig:collageipr} (b), there are several curves colored in purple with an initial non-power law behavior, which will be treated in the following subsection. These coherent states are located across a six-cycle, and as we have already shown in the previous section, their initial non-power-law is due to their closeness to a critical point, in this case, the six-cycle~\cite{Miguel2025}. All the other curves follow the same power law.

\begin{figure}[t]
    \centering
    \includegraphics[width=0.95\columnwidth]{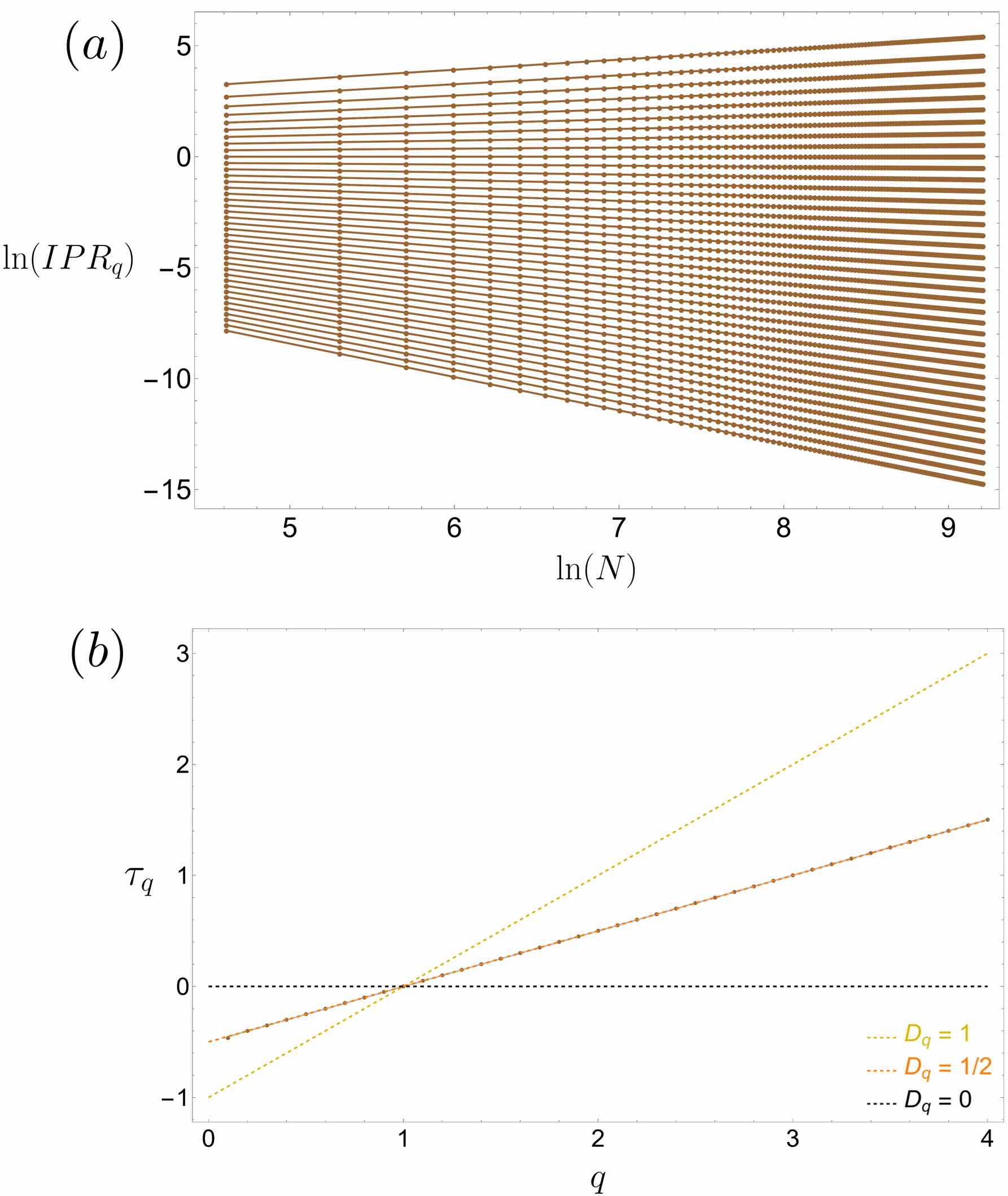}
    \caption{(a) Log-log curve of the $\ln{\text{IPR}_{q}}$ vs $\ln{\left(N\right)}$ of a typical coherent state placed inside a regular part of the phase space of the QKT. From top to bottom, the values of $q$ go from $q=0.1$ to $q=4.0$ in steps of 0.1. (b) $\tau_{q}$ vs $q$ of a coherent state placed inside a regular part of the phase space.}
    \label{fig:10}
\end{figure}

We select the coherent state $|z(0.2,1)\rangle$ as representative among all other coherent states for the cases of $k=0.5$ and $k=1.3$, except the purple ones mentioned in the previous paragraph. Its scaling resembles those of the HO and LMG Hamiltonians. 
In Fig.~\ref{fig:10} (a) it is shown the $\ln{\text{IPR}_{q}}$ vs $\ln{N}$ of this coherent state for forty values of $q$, from 0.1 to 4.0. Notice that these curves refer to a single coherent state. We are not averaging from many states as is a customary procedure in the MFA \cite{Wang2021}. As there is a linear power law scaling, their slopes, the finite mass exponent, are plotted in Fig.~\ref{fig:10} (b). Each brown dot corresponds to the slope of the linear fitting from each curve of Fig.~\ref{fig:10} (a). The other three dashed lines correspond to the three limiting cases explained in Sec.~\ref{sec:3}. The golden dashed line indicates not only the maximally distributed state in a given basis but also corresponds to any state that grows exactly proportionally to the increase of the dimension, as it will be numerically demonstrated in the next section. The third black dashed line corresponds to a state that does not scale with the dimension, which, in our analysis, corresponds to a coherent state located at a critical point inside a regular region. The second orange dashed line corresponds to the analytical $\tau_{q}$ of state having a Gaussian distribution in a given basis ~\cite{Bastarrachea2024}. All of the coherent states analyzed so far, which are not in the immediate vicinity of a critical point, are subject to a formal MFA, and the result is that they acquire the same exact relationship as a Gaussian distribution. Appendix~\ref{app:A} and~\ref{app:B} present the finite mass exponents calculated for all the initial conditions. As mentioned above, more information about the phase space dynamics can be extracted from the $\text{IPR}$ scaling, which will be presented in a future work~\cite{Miguel2025}.

\subsection{Non-generic behavior of the $IPR$ in mixed phase space}

\begin{figure}[ht]
    \centering
    \includegraphics[width=7.5cm,height=5.0cm]{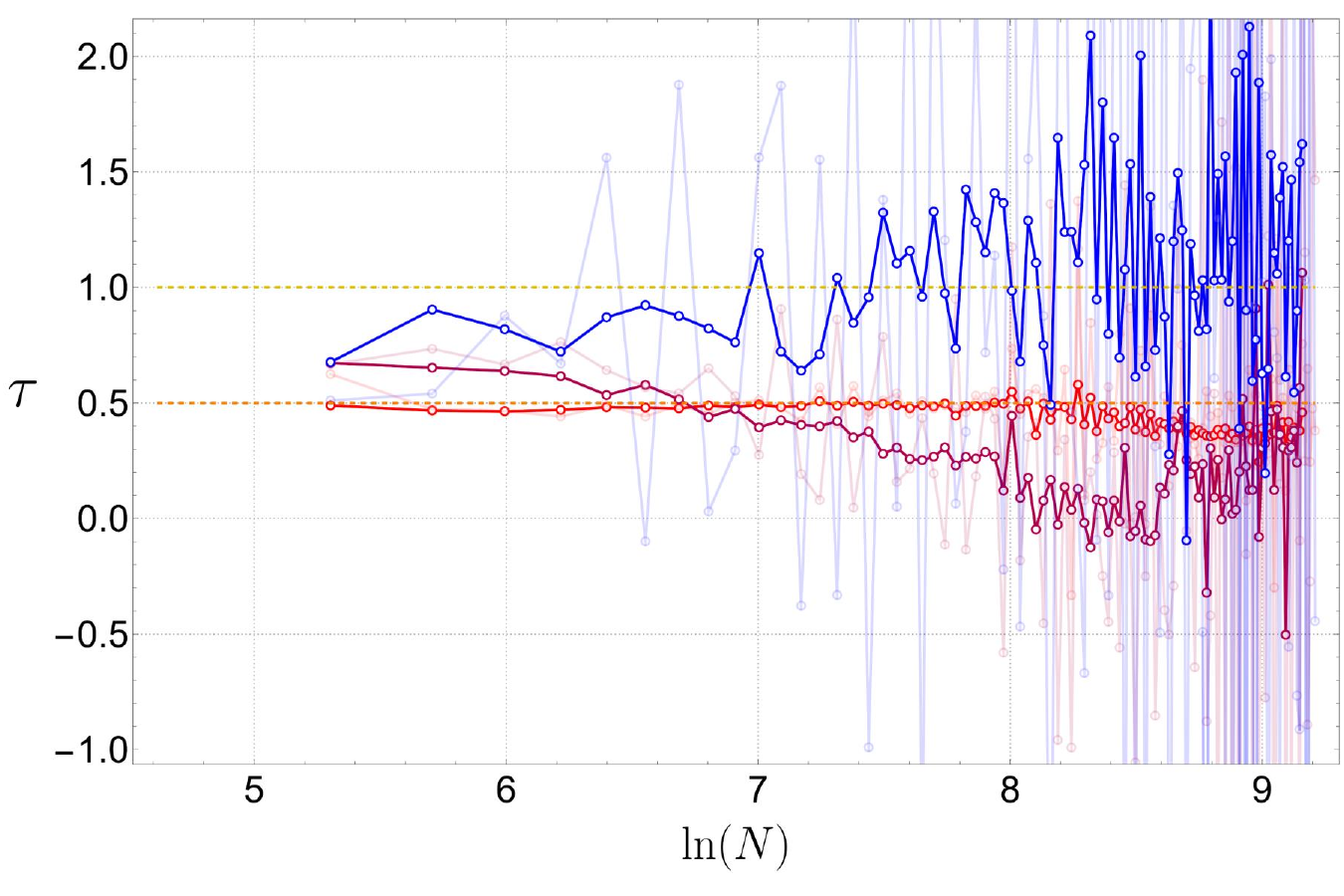}
    \caption{The finite mass exponent of three coherent states with coordinates $P=1$ and $Q=0.2,0.38,0.8$ (red, magenta, and blue respectively) inside mixed phase space of the QKT (see Fig.~\ref{fig:collageipr}). The light tone curves are the real values of $\tau$, and the darker tones are moving averages calculated in intervals of five values. The horizontal dashed lines indicate the asymptotic values for $\tau$ with chaotic (golden) and regular(orange) phase space.}
    \label{fig:iprtypicalmixed}
\end{figure}

The mixed case $k=2.5$ develops chaos in phase space, as seen in the Poincar\'e section in the insert of Fig.~\ref{fig:collageipr}(C). In Fig.~\ref{fig:iprtypicalmixed}, we show the finite mass exponent for three initial conditions in light tones. They display strong fluctuations, which do not get smoother when the dimension is increased. Even taking a moving average, shown in darker tones, the fluctuations remain, making it very difficult to extract information from the scaling of the coherent states. The finite mass exponent calculated for these initial conditions is presented in Appendix~\ref{app:C}.

\subsection{Generic chaotic power-law behavior of the $IPR$}

\begin{figure}[hb]
    \centering
    \includegraphics[width=0.95\columnwidth]{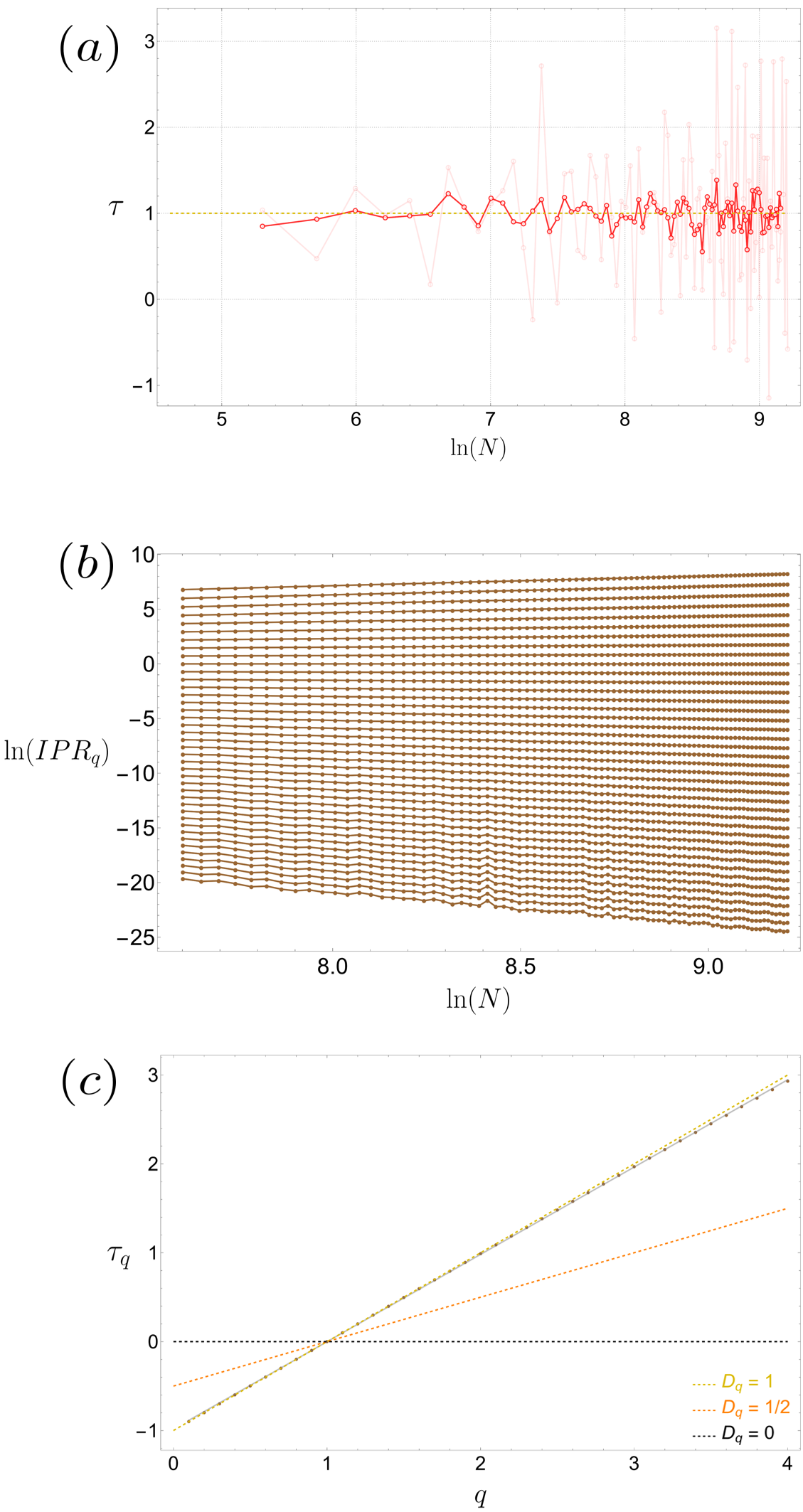}
    \caption{(a) Typical finite mass exponent of the coherent state $|z(0.2,1)\rangle$ for almost chaotic phase space (see Fig.~\ref{fig:collageipr}). The light tone curve is the real value of $\tau$, and the darker tone is a moving average calculated in intervals of five values. The horizontal dashed line indicates the asymptotic value for a typical initial condition inside a chaotic phase space. (b) Log-log curve of the $\ln{\text{IPR}_{q}}$ vs $\ln{\left(N\right)}$ of the coherent state $|z(0.2,1)\rangle$. From top to bottom, the values of $q$ go from $q=0.1$ to $q=4.0$ in steps of 0.1. (c) $\tau_{q}$ vs. $q$ of a typical coherent state placed inside a chaotic part of the phase space.}
    \label{fig:11}
\end{figure}

From the Poincar\'e section in the insert of Fig.~\ref{fig:collageipr}(D), the case $k=30$ can be considered as fully chaotic, despite the possible presence of small regular regions ~\cite{classicaltop1}. In this case, it is possible to perform the MFA for large values of $N$, and the result is a new exponent $\tau=1$ for the power-law, which we identify with the chaotic behavior. It can be observed in Fig, \ref{fig:11} (a), where the finite mass exponent of the coherent state $|z(0.2,1)\rangle$ is shown as a function of $N$ for $k=30$. The slope for each consecutive pair of points in Fig.~\ref{fig:collageipr}(D) is presented with light tone lines, and the darker tone is a moving average of $\tau$ calculated in intervals of five values. The horizontal dashed line indicates the asymptotic value, which is close to 1 and is representative of most of the coherent states for this kicking strength.

In Fig.~\ref{fig:11} (b), we show the $\text{IPR}_{q}$ of the same coherent state $|z(0.2,1)\rangle$, for values of $q \in (0.1,4)$, as a typical example of the other thirty initial conditions. All the lines exhibit a power law scaling, allowing for the MFA. The fractal dimension of all these states is $D_0=1$. In Fig.~\ref{fig:11} (c), we have performed the MFA from $N = 2001$ to $N=10001$ for the chaotic case. Notice that the numerical value for the mass exponent is very close to one. This happens for the other thirty initial conditions, their $D_{q}\approx 1$.

\section{Discussion and Conclusion}
\label{sec:7} 

We have presented a detailed study of the scaling of the inverse participation ratio ($\text{IPR}$) for representative spin coherent states in three different bases, the bounded Harmonic Oscillator, the Lipkin-Meshokov-Glick model, and the Quantum Kicked Top, and related it to its corresponding classical dynamics, including regular, mixed, and fully chaotic regions in phase space. 

We have exhibited that the scaling of the coherent states does not always display a power-law scaling, where a multifractal analysis (MFA) can be performed. To identify the dimensions where MFA is a valuable tool, we introduced a $N$ dependent finite mass exponent $\tau$, which coincides with the standard mass exponent $\tau_{q}$ when, on average, acquires a constant value. This tool has proved to help probe phase space structures, such as regularity, chaos, and fixed points, but it also provides insight into situations where the power-law behavior required to perform a multifractal analysis cannot be asymptotically reached and identify conditions where it emerges. 

Coherent states centered at regions of phase space associated with regularity exhibit, in general, power-law scaling of the $\text{IPR}$ as a function of the dimension $N$, which has the same finite mass exponent $\tau = 1/2 $ for all values $q$ of the mass exponent $\tau_{q}$.  These states have monofractal behavior, with scaling $\text{IPR}_{q} \approx J^{\frac 1 2}$. How fast this asymptotic behavior is reached depends on the closeness of the coherent state to critical points.

The finite mass exponent shows an important consequence of critical points in phase space, i.e., that scaling coherent states close to them needs larger values of $N$ to reach the asymptotic power-law behavior required to perform MFA. This was verified for the integrable rotor associated with a bounded harmonic oscillator, for the LMG model, and the regular regions of the QKT. At the limit of large $N$, the fraction of eigenstates needed to describe any of these coherent states goes to zero, showing that the localization of the coherent states in phase space, whose area scales as $1/J$, goes in parallel with their localization in the Hilbert space. It was also shown that coherent states provide a very accurate description of the eigenstates at fixed points in phase space, implying that their $\text{IPR}$ is close to 1 and independent of the dimension of the Hilbert space. The scaling is  $IPR \approx J^0$, and the mass exponent is null in these cases. Although the calculations presented cover two orders of magnitude of values of $J$, from 50 to 5000, for some states very close to the fixed points, reaching the power-law asymptotic region was not possible. 

On the other hand, it was shown that coherent states in the fully chaotic region, where the Poincar\'e section has no visible structure, the $\text{IPR}$ scales as $\text{IPR} \approx J^1$, with mass exponent 1, and a monofractal behavior. One should underline that this scaling does not imply that any coherent state in the chaotic region has components in all the eigenbasis. It means that the {\em fraction} of eigenstates needed to describe each coherent state remains constant as the dimension of the Hilbert space increases. It was also shown that the scaling of coherent states in the chaotic region of the mixed-phase space has strong fluctuations. In many cases, it was not possible to reach an asymptotic behavior numerically. Again, in these cases, a formal multifractal analysis could be performed.

Studying the scaling of the inverse participation ratio and the finite mass exponents confirms previous results for collective spins systems~\cite{Bastarrachea2016,Wang2021,Wang2023,Bastarrachea2024} but also allows the analysis of individual coherent states, complementary to techniques that rely on averaging over phase space. We expect these results to pave the way to study the mixed phase space and finite-size features of collective spin systems and other algebraic systems with a well-defined classical limit, which are now relevant in the context of quantum technologies. 

\begin{acknowledgments}
We wish to acknowledge S. Lerma-Hernández for his useful comments on these topics. We also acknowledge the support of the Computation Center—ICN, particularly Enrique Palacios, Luciano Díaz, and Eduardo Murrieta. This research was partially funded by DGAPA- UNAM Project No. IN109523.
\end{acknowledgments}

\appendix

\section{Finite mass exponent for $k=0.5$}
\label{app:A}

\begin{figure}[h]
    \centering
    \includegraphics[width=\columnwidth]{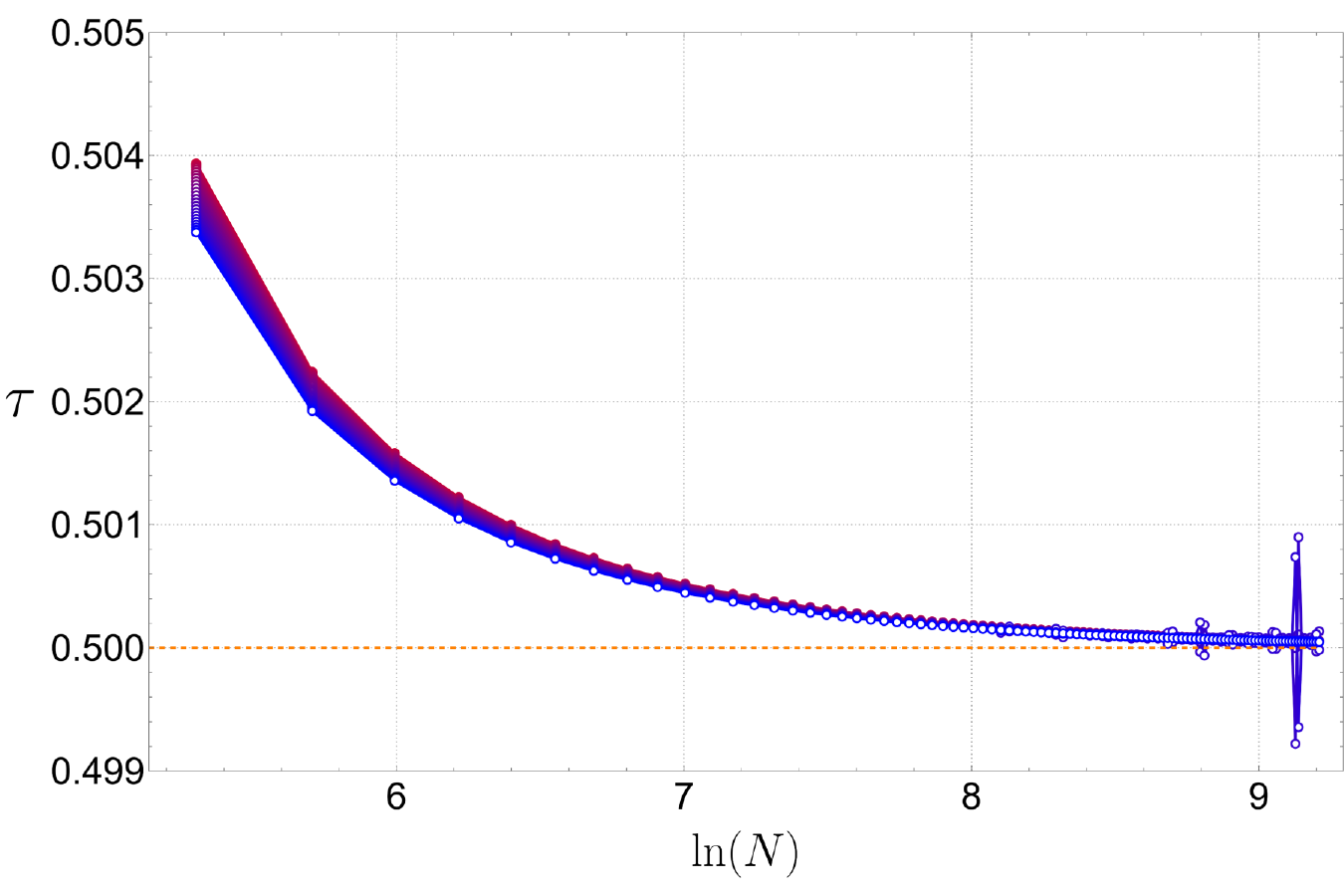}
    \caption{The finite mass exponent of the thirty-one initial conditions with coordinates $P=1$ and $Q\in [0.2,0.8]$ of the QKT.}
    \label{fig:alltauregular}
\end{figure}

In Fig.~\ref{fig:alltauregular}, we show all the finite mass exponent calculated for the thirty-one initial conditions with coordinates $P=1$ and $Q\in [0.2,0.8]$ for the QKT. Notice that the value of all the $\tau$'s converge quickly to the expected value obtained 
in the previous sections, allowing the MFA to be performed. In the end, near $\ln(N)=9$, there is a disturbance in the value of the finite mass exponent for some initial conditions. It claims for a more detailed analysis, which is in process and will be presented in future article~\cite{Miguel2025}.

\section{Finite mass exponent for $k=1.3$}
\label{app:B}

\begin{figure}[h]
    \centering
    \includegraphics[width=0.9\columnwidth]{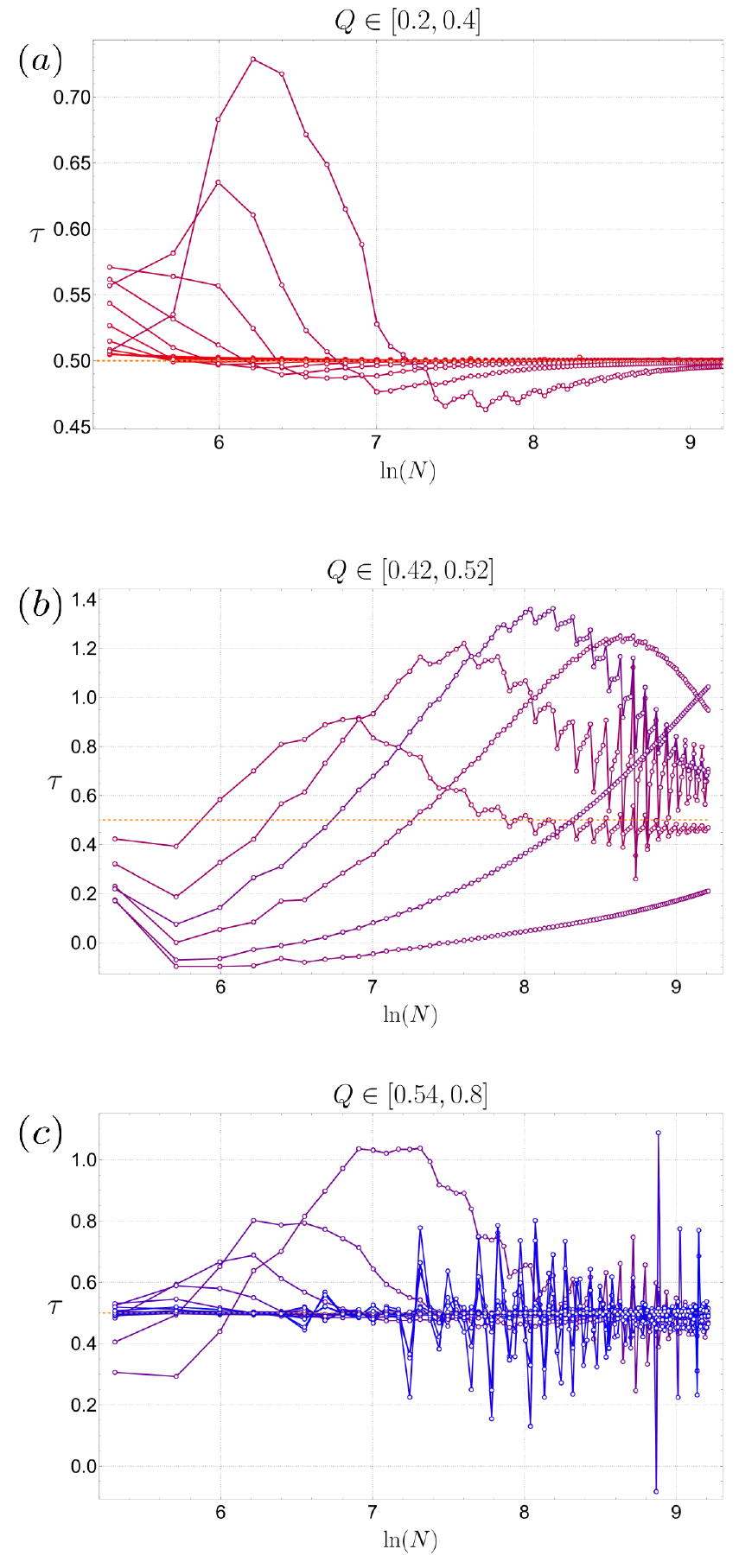}
    \caption{(a) Finite mass exponent for the coherent states with coordinates at (a) $P=1$ and $Q\in [0.2,0.4]$ (b) $P=1$ and $Q\in [0.42,0.52]$, and (c) $P=1$ and $Q\in [0.54,0.8]$ of the QKT (same color code as the Fig.~\ref{fig:collageipr}). The horizontal dashed line indicates the asymptotic values for $\tau$ inside regular phase space.}
    \label{fig:A2}
\end{figure}

All the other mass exponents for the case of $k=1.3$ of the QKT are presented. Although there are regular dynamics in all the phase space, there are several new critical points compared with the LMG model due to the breaking of integrability induced by the kicking potential. This new geometry of the phase space is also captured by the scaling of the $\text{IPR}$. For the initial conditions analyzed in the main text (\ref{sec:6}) where it was possible to perform the MFA, in Figs.~\ref{fig:A2} (a) and~\ref{fig:A2} (b), the finite mass exponent are shown, exhibiting the result for regular phase space $D_0 = \frac{1}{2}$. In Fig.~\ref{fig:A2} (c), we show the initial conditions where the MFA cannot be performed due to the closeness of those coherent states to the center of that ellipse (see Fig.~\ref{fig:collageipr}), which belongs to a six-cycle.

\section{Closer look to mixed phase space for $k=2.5$}
\label{app:C}

\begin{figure*}[ht]
    \centering
    \includegraphics[width=\textwidth]{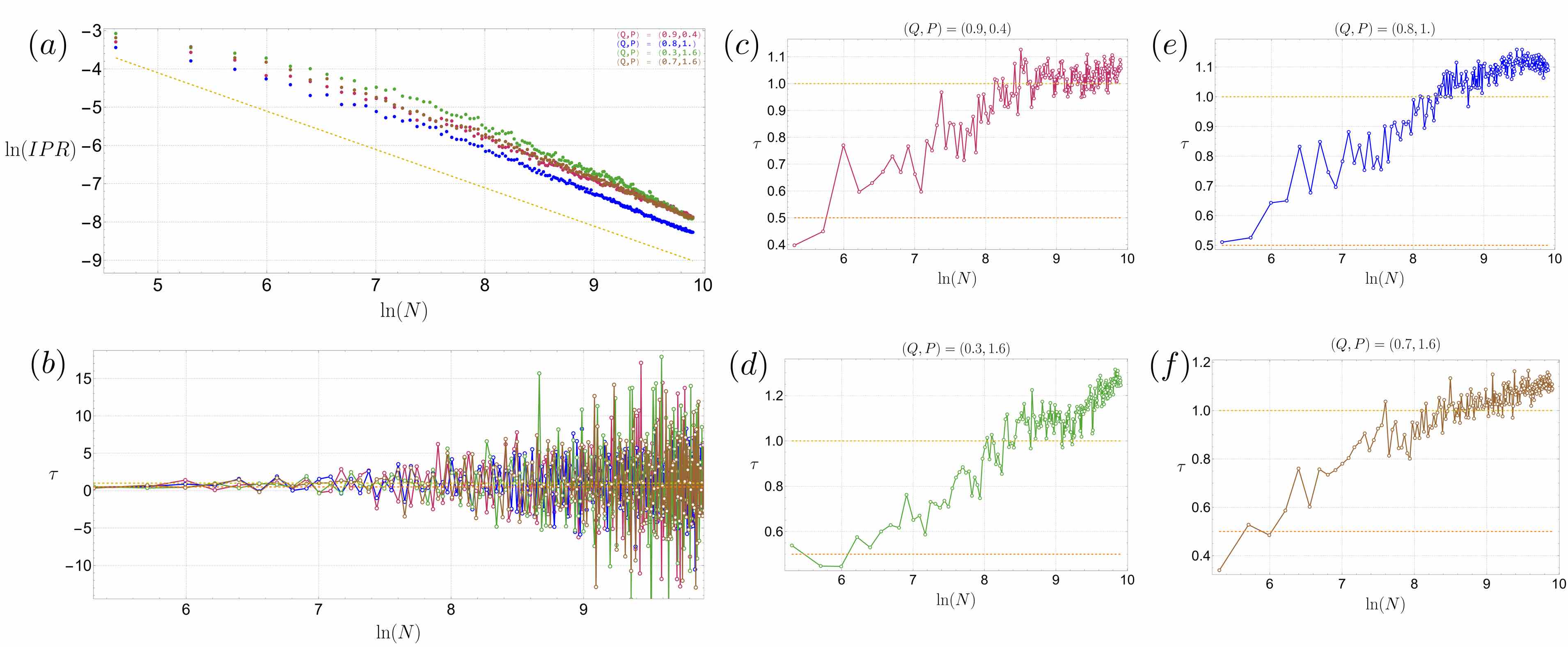}
    \caption{(a) Scaling of $\text{IPR}$ for four initial conditions completely inside of the chaotic part of mixed phase space of the QKT (see Fig.~\ref{fig:collageipr}). The golden dashed line is a straight line with a slope of one, just for reference. (b) Finite mass exponent for four initial conditions completely inside the chaotic part of mixed phase space (see Fig.~\ref{fig:collageipr}). The horizontal golden(orange) dashed line corresponds to the asymptotic value found for chaotic(regular) dynamics. (c)-(f) Cumulative finite mass exponent for four initial conditions completely inside of the chaotic part of mixed phase space (see Fig.~\ref{fig:collageipr}). The horizontal golden(orange) dashed line corresponds to the asymptotic value found for chaotic(regular) dynamics.}
    \label{fig:A3}
\end{figure*}

Here, we 
show the $\text{IPR}$ of another four initial conditions placed inside the chaotic part of the mixed phase space of the QKT. This time, we scaled the system from $J=50$ up to $J=10050$ (dimension $N=20101$) in steps of $\Delta J=50$ in order to see if the finite mass exponent would stabilize its value so the MFA could be performed. We show the scaling in Fig.~\ref{fig:A3} (a). The finite mass exponent gives the result shown in Fig.~\ref{fig:A3} (b). As can be seen in this figure, their value fluctuates very erratically. To be able to see in greater detail around what value those finite mass exponents are fluctuating, in Fig.~\ref{fig:A3} (c)-(f), we have calculated the cumulative sum as a function of the dimension of the values shown in Fig.~\ref{fig:A3} (b). Three curves tend to fluctuate around the value one, which we have related to the chaotic dynamics. For the initial condition plotted in green one there is not a clear limit.

\bibliography{Bibliography}

\end{document}